\documentclass[twocolumn,preprintnumbers,amsmath,amssymb]{revtex4}

\usepackage{graphicx}
\usepackage{dcolumn}
\usepackage{bm}

\newcommand{\bra}[1]{\ensuremath{\langle #1 \vert}}
\newcommand{\ket}[1]{\ensuremath{\vert #1  \rangle}}
\newcommand{\braket}[2]{\ensuremath{\langle  #1 \vert #2  \rangle}}

\renewcommand{\b}[1]{\ensuremath{\mathbf{#1}}}
\newcommand{\VMC}{\ensuremath{\text{VMC}}}
\newcommand{\DMC}{\ensuremath{\text{DMC}}}
\newcommand{\CSF}{\ensuremath{\text{CSF}}}

\newcommand{\EFP}{\ensuremath{\text{EFP}}}

\renewcommand{\L}{\ensuremath{\text{L}}}
\newcommand{\diag}{\ensuremath{\text{diag}}}
\newcommand{\bas}{\ensuremath{\text{bas}}}
\newcommand{\orb}{\ensuremath{\text{orb}}}
\newcommand{\opt}{\ensuremath{\text{opt}}}
\newcommand{\Jas}{\ensuremath{\text{Jas}}}
\newcommand{\lin}{\ensuremath{\text{lin}}}
\newcommand{\Psit}{\ensuremath{\tilde{\Psi}}}
\newcommand{\Psib}{\ensuremath{\overline{\Psi}}}
\newcommand{\Psibb}{\ensuremath{\overline{\overline{\Psi}}}}

\newcommand{\vac}{\ensuremath{\text{vac}}}

\newcommand{\exact}{\ensuremath{\text{exact}}}
\newcommand{\pert}{\ensuremath{\text{pert}}}
\renewcommand{\l}{\ensuremath{\lambda}}

\newcommand{\pb}{\ensuremath{\overline{p}}}
\newcommand{\pbb}{\ensuremath{\overline{\overline{p}}}}

\def\Hhat{\hat{H}}

\begin{document}

\title{Optimization of quantum Monte Carlo wave functions by energy minimization}

\author{Julien Toulouse}
\email{toulouse@tc.cornell.edu}
\affiliation{
Cornell Theory Center, Cornell University, Ithaca, New York 14853, USA.
}
\author{C. J. Umrigar}
\email{cyrus@tc.cornell.edu}
\affiliation{
Cornell Theory Center and Laboratory of Atomic and Solid State Physics, Cornell University, Ithaca, New York 14853, USA.
}

\date{\today}

\begin{abstract}
We study three wave function optimization methods based on energy minimization in a variational Monte Carlo framework: the Newton, linear and perturbative methods. In the Newton method, the parameter variations are calculated from the energy gradient and Hessian, using a reduced variance statistical estimator for the latter. In the linear method, the parameter variations are found by diagonalizing a nonsymmetric estimator of the Hamiltonian matrix in the space spanned by the wave function and its derivatives with respect to the parameters, making use of a strong zero-variance principle. In the less computationally expensive perturbative method, the parameter variations are calculated by approximately solving the generalized eigenvalue equation of the linear method by a nonorthogonal perturbation theory. These general methods are illustrated here by the optimization of wave functions consisting of a Jastrow factor multiplied by an expansion in configuration state functions (CSFs) for the C$_2$ molecule, including both valence and core electrons in the calculation. The Newton and linear methods are very efficient for the optimization of the Jastrow, CSF and orbital parameters. The perturbative method is a good alternative for the optimization of just the CSF and orbital parameters. Although the optimization is performed at the variational Monte Carlo level, we observe for the C$_2$ molecule studied here, and for other systems we have studied, that as more parameters in the trial wave functions are optimized, the diffusion Monte Carlo total energy improves monotonically, implying that the nodal hypersurface also improves monotonically.
\end{abstract}

\maketitle

\section{Introduction}
\label{sec:intro}

Quantum Monte Carlo (QMC) methods (see e.g.
Refs.~\onlinecite{HamLesRey-BOOK-94,NigUmr-BOOK-99,FouMitNeeRaj-RMP-01})
constitute an alternative to standard {\it ab initio} methods of
quantum chemistry for accurate calculations of the electronic
structure of atoms, molecules and solids. The two most commonly used
variants, variational Monte Carlo (VMC) and diffusion Monte Carlo
(DMC), rely on an explicitly correlated trial wave function, generally
consisting for atoms and molecules of a Jastrow factor multiplied by a short expansion in
configuration state functions (CSFs), each consisting of a linear
combination of Slater determinants, a
form capable of encompassing most of the electron correlation effects.
To fully benefit from the considerable flexibility in the form of the
wave function, it is crucial to be able to efficiently optimize the
parameters in these wave functions.

Variance minimization in correlated
sampling~\cite{UmrWilWil-PRL-88,UmrWilWil-INC-88,Umr-IJQC-89} has
become the most frequently used method in QMC for optimizing wave
functions because it is far more efficient than {\it straightforward}
energy minimization on a finite Monte Carlo sample. However, while the method works relatively well
for the optimization of the Jastrow factor, it is much less effective
for the optimization of the determinantal part of the wave function
(though still
possible~\cite{UmrWilWil-PRL-88,FilUmr-JCP-96,HuaUmrNig-JCP-97}).
Further, there is some evidence that energy-optimized wave
functions give on average better expectation values for other
observables than variance-optimized ones (see, e.g.,
Refs.~\onlinecite{SnaRot-JCP-00,GalBueSar-JCP-01}). As a result, a lot
of effort has recently been devoted to developing efficient
methods for the optimization of QMC wave functions by energy minimization.
On the other hand, it should be mentioned that variance-minimized wave
functions often have a smaller time-step error in DMC.

We now summarize some of the major approaches that have been proposed for energy
minimization in VMC. The most efficient method to
minimize the energy with respect to linear parameters, such as the CSF
coefficients, is to solve the associated generalized eigenvalue
equation using a non-symmetric estimator of the Hamiltonian matrix~\cite{NigMel-PRL-01}.
The energy fluctuation potential (EFP)
method~\cite{Fah-INC-99,FilFah-JCP-00,PreBevFah-PRB-02,SchFah-JCP-02,SchFil-JCP-04}
is very efficient for optimizing some nonlinear parameters and has
been applied very successfully to the optimization of the
orbitals~\cite{FilFah-JCP-00,SchFil-JCP-04} and CSF
coefficients~\cite{SchFah-JCP-02,SchFil-JCP-04}. It has also been
applied to the optimization of Jastrow factors in periodic
solids~\cite{PreBevFah-PRB-02}. The perturbative EFP method, a
simplification of the EFP method, retains the same convergence rate
for the optimization of the orbitals and CSF coefficients while
decreasing the computational cost~\cite{SceFil-PRB-06}. The stochastic
reconfiguration (SR) method, originally developed for lattice
systems~\cite{Sor-PRB-01}, has been applied to the full optimization
of atomic and molecular wave functions consisting of an
antisymmetrized geminal power part multiplied by a Jastrow
factor~\cite{CasSor-JCP-03,CasAttSor-JCP-04}. It is related to the
perturbative EFP method and is simpler but less
efficient~\cite{SchFil-JCP-04,SceFil-PRB-06}. The Newton
method is a conceptually simple and general optimization method but a straightforward implementation
of it in QMC is rather inefficient~\cite{LinZhaRap-JCP-00,LeeMelRap-JCP-05}.
However, an improved version of it, making use of a reduced variance
estimator of the Hessian matrix~\cite{UmrFil-PRL-05}, is very efficient
for the optimization of Jastrow factors. Another modified version of the Newton method with an
approximate Hessian, named stochastic reconfiguration with Hessian acceleration (SRH), has been applied to lattice models~\cite{Sor-PRB-05}.

In this work, we investigate the three best energy minimization methods for the
optimization of the Jastrow, CSF and orbital parameters of QMC wave
functions: the Newton, linear and
perturbative methods. The Newton method has already been
applied very successfully to the optimization of Jastrow factors by
Umrigar and Filippi~\cite{UmrFil-PRL-05}, and in this paper it is
also applied to the optimization of the determinantal part of the wave
function. The linear method is an extension
of the zero-variance generalized eigenvalue equation approach of Nightingale and
Melik-Alaverdian~\cite{NigMel-PRL-01} to arbitrary nonlinear
parameters: at each step of the iterative procedure, the wave function
is linearized with respect to the parameters and the optimal values of
the parameters are found by diagonalizing the Hamiltonian in the space
spanned by the current wave function and its derivatives with respect to the parameters.
This method is briefly presented in Ref.~\onlinecite{UmrTouFilSorHeg-JJJ-XX}.
The perturbative method coincides with the perturbative EFP method of
Scemama and Filippi~\cite{SceFil-PRB-06} for the optimization of the
CSF and orbital parameters. Here, we put this approach on more general
grounds by recasting it as a simplification of the linear method where
the generalized eigenvalue equation is solved approximately by a
nonorthogonal perturbation theory. The Newton and linear methods are
very efficient for the optimization of the Jastrow, CSF and orbital
parameters. The perturbative method is a good alternative for the
optimization of just the CSF and orbital parameters.

The paper is organized as follows. In Sec.~\ref{sec:wfparam}, the
parametrization of the trial wave function is presented. The
energy minimization procedures are discussed in Sec.~\ref{sec:emin},
and their realizations in VMC are discussed in Sec~\ref{sec:vmc}.
Sec.~\ref{sec:compdetails} contains computational details of the
calculations performed on the C$_2$ molecule to test the optimization
methods, and in Sec.~\ref{sec:results} we present the
results. Sec.~\ref{sec:conclusion} contains our conclusions.

Hartree atomic units (Ha) are used throughout this work.

\section{Wave function parametrization and derivatives}
\label{sec:wfparam}

We begin by describing the form of the wave function used, the actual
parametrization chosen for the optimization, and the corresponding
derivatives of the wave function with respect to the parameters.

\subsection{Form of the wave function}

We use an $N$-electron wave function of the usual Jastrow-Slater form
that is denoted at each iteration of the optimization procedure by
\begin{eqnarray}
\ket{\Psi_0} = \hat{J}(\bm{\alpha}^0) \ket{\Phi_0},
\label{Psi0}
\end{eqnarray}
where $\hat{J}(\bm{\alpha}^0)$ is a Jastrow operator depending on the
current parameters $\alpha_i^0$ and $\ket{\Phi_0}$ is a
multi-determinantal wave function. For notational convenience, we assume that the
wave function $\ket{\Psi_0}$ is always normalized to unity, i.e.
$\braket{\Psi_0}{\Psi_0}=1$. In practice, $\ket{\Psi_0}$ can have arbitrary normalization.

The wave function $\ket{\Phi_0}$ is a linear combination of $N_{\CSF}$
orthonormal configuration state functions (CSFs), $\ket{C_{I}}$, with
current coefficients $c_{I}^{0}$,
\begin{eqnarray}
\ket{\Phi_0} =  \sum_{I=1}^{N_{\CSF}} c_{I}^{0} \ket{C_{I}}.
\end{eqnarray}
Each CSF is a short linear combination of products of spin-up and
spin-down Slater determinants, $\ket{D_{\b{k}}^\uparrow}$ and
$\ket{D_{\b{k}}^\downarrow}$, $\ket{C_{I}} = \sum_{\b{k}} d_{I,\b{k}}
\ket{D_{\b{k}}^{\uparrow}} \ket{D_{\b{k}}^{\downarrow}}$, where the coefficients $d_{I,\b{k}}$ are fully
determined by the spatial and spin symmetries of the state considered
(see, e.g., Ref.~\onlinecite{HuaFilUmr-JCP-98}). The use of CSFs is
important to decrease the number of coefficients to be optimized and
to ensure the correct symmetry of the wave function after optimization
in the presence of statistical noise. The $N_{\uparrow}$-electron and
$N_{\downarrow}$-electron spin-assigned Slater determinants are
generated from a set of current orthonormal orbitals,
$\ket{D_{\b{k}}^\uparrow} = \hat{a}^{\dag}_{k_1\uparrow}
\hat{a}^{\dag}_{k_2\uparrow} \cdots \, \hat{a}^{\dag}_{k_{N_\uparrow}
  \uparrow} \ket{\vac}$ and $\ket{D_{\b{k}}^\downarrow} =
\hat{a}^{\dag}_{k_{N_\uparrow+1} \downarrow}
\hat{a}^{\dag}_{k_{N_\uparrow+2} \downarrow} \cdots \,
\hat{a}^{\dag}_{k_N \downarrow} \ket{\vac}$, where $\hat{a}^{\dag}_{k
  \, \sigma}$ (with $\sigma=\uparrow, \downarrow$) is the fermionic
creation operator for the spatial orbital $\ket{\phi_{k}^0}$ in the
spin-$\sigma$ determinant, and $\ket{\vac}$ is the vacuum state of
second quantization. The (occupied and virtual) spatial orbitals are
written as linear combinations of $N_{\bas}$ basis functions
$\ket{\chi_{\mu}}$ (e.g., Slater or Gaussian functions) with current
coefficients $\lambda_{k,\mu}^{0}$, $\ket{\phi_{k}^0} =
\sum_{\mu=1}^{N_{\bas}} \lambda_{k,\mu}^{0} \ket{\chi_{\mu}}$.

The $N$-electron Jastrow operator, $\hat{J}(\bm{\alpha}^0)$, is defined
by its matrix elements in the $N$-electron position basis
$\ket{\b{R}}=\ket{\b{r}_1,\b{r}_2,...,\b{r}_N}$
\begin{eqnarray}
\bra{\b{R}} \hat{J}(\bm{\alpha}^0) \ket{\b{R}'} = J(\bm{\alpha}^0;\b{R}) \delta(\b{R}-\b{R}'),
\end{eqnarray}
where $J(\bm{\alpha}^0;\b{R})$ is the spin-assigned Jastrow factor, a
real positive function of $\b{R}$ which is symmetric under the exchange of two same-spin electrons. Its action on an arbitrary
$N$-electron state $\ket{\Phi}$ is given by $\bra{\b{R}}
\hat{J}(\bm{\alpha}^0) \ket{\Phi} = J(\bm{\alpha}^0;\b{R}) \Phi(\b{R}) $
where $\Phi(\b{R}) = \braket{\b{R}}{\Phi}$. The Jastrow operator is
Hermitian, $\hat{J}(\bm{\alpha}^0)^{\dag} =
\hat{J}(\bm{\alpha}^0)$. We use flexible Jastrow factors consisting of the
exponential of the sum of electron-nucleus, electron-electron
and electron-electron-nucleus terms, written as systematic
polynomial or Pad\'e expansions~\cite{Umr-UNP-XX} (see also
Refs.~\onlinecite{FilUmr-JCP-96,GucSanUmrJai-PRB-05}).

\subsection{Wave function parametrization}

We want to optimize the Jastrow parameters $\alpha_i$, the CSF
coefficients $c_I$ and the orbital coefficients $\lambda_{k,\mu}$. Some parameters in the Jastrow factor
are fixed by imposing the electron-nucleus and electron-electron cusp conditions~\cite{Kat-CPAM-57}
on the wave function; the other Jastrow parameters are varied freely.
Due to the arbitrariness of
the overall normalization of the wave function, only $N_{\CSF}-1$ CSF
coefficients need be varied, e.g., the coefficient of the first
configuration can be kept fixed. The situation is more involved for
the orbital coefficients which are not independent
due to the invariance properties of determinants under elementary row operations.
To easily retain only
unconstrained, nonredundant orbital parameters in the optimization, it is convenient to
vary the orbital coefficients by performing
rotations among the (occupied and virtual) orbitals with a unitary operator
parametrized as an exponential of an anti-Hermitian operator.
This parametrization is
used in multi-configuration self-consistent-field (MCSCF)
calculations (for a recent and general review of MCSCF theory, see
Ref.~\onlinecite{SchGor-ARPC-98}). More specifically, we use the
following parametrization of the wave function depending on $N^\opt_\Jas$ Jastrow parameters $\bm{\alpha}$,
$N^\opt_\CSF = N_{\CSF}-1$ free CSF coefficients $\b{c}$ ($c_1$ is fixed), and $N^\opt_\orb$ orbital rotation parameters $\bm{\kappa}$
\begin{eqnarray}
\ket{\Psi(\bm{\alpha},\b{c},\bm{\kappa})} =  \hat{J}(\bm{\alpha}) e^{\hat{\kappa}(\bm{\kappa})} \sum_{I=1}^{N_\CSF} c_I \ket{C_I},
\label{Psi}
\end{eqnarray}
where $e^{\hat{\kappa}(\bm{\kappa})}$ is the unitary operator that
performs rotations in orbital space (see, e.g., Refs.~\onlinecite{Jen-INC-94,HelJorOls-BOOK-02}).
More elaborate parametrizations of the CSF coefficients, such as a unitary
parametrization~\cite{Dal-CPL-79}, are often used in MCSCF theory
(see, e.g., Ref.~\onlinecite{HelJorOls-BOOK-02}), but we have not found any
decisive advantage to using them for our purpose.

The rotations in orbital space are generated by the anti-Hermitian
real singlet orbital excitation operator~\cite{DalJor-JCP-78}
\begin{eqnarray}
\hat{\kappa}(\bm{\kappa}) = \sum_{k<l} \kappa_{kl} \,  \hat{E}_{kl}^{-},
\label{}
\end{eqnarray}
where the sum is over all nonredundant orbital pairs,
$\hat{E}_{kl}^{-} = \hat{E}_{kl} - \hat{E}_{lk}$, and $\hat{E}_{kl} =
\hat{a}_{k \uparrow}^{\dag} \hat{a}_{l \uparrow} + \hat{a}_{k
  \downarrow}^{\dag} \hat{a}_{l \downarrow} $ is the singlet
excitation operator from orbital $l$ to orbital $k$. In
Eq.~(\ref{Psi}), the action of the operator
$e^{\hat{\kappa}(\bm{\kappa})}$ is to rotate each occupied orbital in
the Slater determinants as
\begin{eqnarray}
\ket{\phi_k} = e^{\hat{\kappa}(\bm{\kappa})} \ket{\phi_k^{0}} = \sum_{l} (e^{\bm{\kappa}})_{kl} \ket{\phi_l^{0}},
\label{phik}
\end{eqnarray}
where the sum is over all (occupied and virtual) orbitals, and
$(e^{\bm{\kappa}})_{kl}$ are the elements of the orthogonal matrix
$e^{\bm{\kappa}}$ constructed from the real anti-symmetric matrix
$\bm{\kappa}$ with elements $\kappa_{kl}$. More generally, any unitary matrix can be written as an exponential of an anti-Hermitian matrix, the off-diagonal upper triangular part of the anti-Hermitian matrix realizing a nonredundant parametrization of the unitary matrix. To maintain the orthonormality of
the entire set of orbitals, the operator
$e^{\hat{\kappa}(\bm{\kappa})}$ is applied to the virtual orbitals as well.
For a single Slater determinant wave function, the orbitals can
be partitioned into three sets referred to as \textit{closed} (i.e.,
doubly occupied), \textit{open} (i.e., singly occupied) and
\textit{virtual} (i.e., unoccupied). The nonredundant excitations to
consider are then: closed $\to$ open, closed $\to$ virtual and open
$\to$ virtual. For a multiconfiguration wave function, the orbitals
can be partitioned into three sets referred to as \textit{inactive}
(i.e., occupied in all determinants), \textit{active} (i.e., occupied
in some determinants and unoccupied in the others) and
\textit{secondary} (i.e., unoccupied in all determinants). For a
multiconfiguration complete active space (CAS) wave
function~\cite{RooTaySie-CP-80}, the nonredundant excitations are
then: inactive $\to$ active, inactive $\to$ secondary and active $\to$
secondary. For a single-determinant and multi-determinant CAS wave function,
the action of the reverse excitation from orbital $k$ to $l$ ($\hat{E}_{lk}$) in
$\hat{E}_{kl}^{-}=\hat{E}_{kl} - \hat{E}_{lk}$ is always zero. For a general
multiconfiguration wave function (not CAS), some active $\to$ active excitations
must also be included. Consequently, the action of the reverse
excitation $\hat{E}_{lk}$ in $\hat{E}_{kl}^{-}=\hat{E}_{kl} - \hat{E}_{lk}$
does not generally vanish. Only excitations between orbitals
of the same spatial symmetry have to be considered.
In the super configuration interaction approach~\cite{GreCha-CPL-71} where the
orbitals are optimized by adding the single excitations of the
(multiconfiguration) reference wave function to the variational
space, pioneered in QMC by Filippi and
coworkers~\cite{SchFil-JCP-04,SceFil-PRB-06},
an alternative linear parametrization of the orbital space is chosen,
$\ket{\phi_k} = \left( \hat{1} + \hat{\kappa}(\bm{\kappa}) \right) \ket{\phi_k^{0}}$,
instead of the unitary parametrization of Eq.~(\ref{phik}). In that case, the
optimized orbitals are not orthonormal.

In the following, we will collectively refer to the Jastrow, CSF and orbital parameters as $\b{p}=(\bm{\alpha},\b{c},\bm{\kappa})$. The wave function of Eq.~(\ref{Psi0}) is thus simply $\ket{\Psi_0} = \ket{\Psi(\b{p}^0)}$ where $\b{p}^0=(\bm{\alpha}^0,\b{c}^0,\bm{\kappa}^0=\bm{0})$ are the current parameters. We will designate by $N^\opt = N^\opt_\Jas + N^\opt_\CSF + N^\opt_\orb$ the total number of parameters to be optimized.

\subsection{First-order wave function derivatives}
\label{sec:deriv1}

We now give the expressions for the first-order derivatives of the wave
function $\ket{\Psi(\b{p})}$ of Eq.~(\ref{Psi}) with respect to
the parameters $p_i$ at $\b{p}=\b{p}^0$
\begin{eqnarray}
\ket{\Psi_i} = \left( \frac{\partial \ket{\Psi(\b{p})}}{\partial p_i} \right)_{\b{p}=\b{p}^0},
\label{Psii}
\end{eqnarray}
which collectively designate the derivatives with respect to the Jastrow parameters
\begin{eqnarray}
\ket{\Psi_{\alpha_i}}= \frac{\partial \hat{J}(\bm{\alpha}^0)}{\partial \alpha_i} \ket{\Phi_0},
\label{}
\end{eqnarray}
with respect to the CSF parameters
\begin{eqnarray}
\ket{\Psi_{c_I}} &=&  \hat{J}(\bm{\alpha}^0) \ket{C_I},
\label{}
\end{eqnarray}
and with respect to the orbital parameters
\begin{eqnarray}
\ket{\Psi_{\kappa_{kl}}} &=& \hat{J}(\bm{\alpha}^0) \hat{E}_{kl}^{-} \ket{\Phi_0}.
\label{}
\end{eqnarray}
The first-order orbital derivatives are thus generated by the single
excitations of orbitals out of the state $\ket{\Phi_0}$.

\subsection{Second-order wave function derivatives}

The second-order derivatives
with respect to the parameters $p_i$ at $\b{p}=\b{p}^0$,
which are needed only for the Newton method, are
\begin{eqnarray}
\ket{\Psi_{ij}}= \left( \frac{\partial^2 \ket{\Psi(\b{p})}}{\partial p_i \partial p_j} \right)_{\b{p}=\b{p}^0},
\label{Psiij}
\end{eqnarray}
which collectively designate the Jastrow-Jastrow derivatives
\begin{eqnarray}
\ket{\Psi_{\alpha_i\alpha_j}}= \frac{\partial^2 \hat{J}(\bm{\alpha}^0)}{\partial \alpha_i \partial \alpha_j} \ket{\Phi_0},
\label{}
\end{eqnarray}
the Jastrow-CSF derivatives
\begin{eqnarray}
\ket{\Psi_{\alpha_i c_I}}= \frac{\partial \hat{J}(\bm{\alpha}^0)}{\partial \alpha_i} \ket{C_I},
\label{}
\end{eqnarray}
the Jastrow-orbital derivatives
\begin{eqnarray}
\ket{\Psi_{\alpha_i \kappa_{kl}}}= \frac{\partial \hat{J}(\bm{\alpha}^0)}{\partial \alpha_i} \hat{E}_{kl}^{-} \ket{\Phi_0},
\label{}
\end{eqnarray}
the CSF-orbital derivatives
\begin{eqnarray}
\ket{\Psi_{c_I \kappa_{kl}}} &=&  \hat{J}(\bm{\alpha}^0) \hat{E}_{kl}^{-} \ket{C_I},
\label{}
\end{eqnarray}
and the orbital-orbital derivatives
\begin{eqnarray}
\ket{\Psi_{\kappa_{kl} \kappa_{mn}}} &=& \hat{J}(\bm{\alpha}^0) \hat{E}_{kl}^{-} \hat{E}_{mn}^{-} \ket{\Phi_0}.
\label{}
\end{eqnarray}
Notice that the wave function form of Eq.~(\ref{Psi}) is linear in the
CSF parameters and therefore the CSF-CSF derivatives are zero,
$\ket{\Psi_{c_I c_J}} = 0$. The orbital-orbital derivatives correspond
to double excitations of orbitals out of the state $\ket{\Phi_0}$.
Since we usually start the optimization with reasonably good initial
orbitals coming from a standard MCSCF calculation we set these
second derivatives to zero, $\ket{\Psi_{\kappa_{kl} \kappa_{mn}}} = 0$, in order
to reduce the computational cost per iteration during Newton minimization.
Nevertheless, it takes only a few steps to optimize the orbitals
as discussed in Sec.~\ref{sec:results}.

\section{Energy minimization procedures}
\label{sec:emin}

In this section, we present the three methods investigated in this
work to minimize the variational energy with respect to the wave
function parameters $\b{p}$
\begin{eqnarray}
E = \min_{\b{p}} E(\b{p}),
\label{Evar}
\end{eqnarray}
where $E(\b{p}) = \bra{\Psi(\b{p})} \hat{H} \ket{\Psi(\b{p})} /\braket{\Psi(\b{p})}{\Psi(\b{p})}$ and $\hat{H} =\hat{T} + \hat{W}_{ee} + \hat{V}_{ne}$
is the electronic Hamiltonian, including the kinetic, electron-electron interaction
and nuclei-electron interaction terms. The Hamiltonian can also include a
nonlocal pseudopotential, enabling one to avoid the explicit treatment of core electrons.
The energy corresponding to the current parameters $\b{p}^0$ will be denoted by
$E_0 = E(\b{p}^0)$.

\subsection{Newton optimization method}

The Newton method was first applied to the optimization of QMC wave functions by Rappe and
coworkers~\cite{LinZhaRap-JCP-00,LeeMelRap-JCP-05}. It has been
considerably improved by Umrigar and Filippi~\cite{UmrFil-PRL-05}, and
by Sorella~\cite{Sor-PRB-05}, by making use of a lower variance statistical
estimator of the Hessian matrix and by employing stabilization techniques.
In Ref.~\onlinecite{UmrFil-PRL-05} the correct Hessian was used, whereas in
Ref.~\onlinecite{Sor-PRB-05} an approximate Hessian, which reduces to the exact
Hessian for parameters that are linear in the exponent, was used.
We now recall the basic working equations.

The energy $E(\b{p})$ is expanded to second-order in the
parameters $\b{p}$ around $\b{p}^0$
\begin{eqnarray}
E^{[2]} (\b{p}) = E_0 + \sum_{i=1}^{N^\opt} g_i \Delta p_i + \frac{1}{2} \sum_{i=1}^{N^\opt} \sum_{j=1}^{N^\opt} h_{ij} \Delta p_i \Delta p_j,
\nonumber\\
\label{E2}
\end{eqnarray}
where the sums are over all the parameters to be optimized, $\Delta p_i = p_i - p_i^0$
are the components of the vector of parameter variations $\Delta \b{p}$,
\begin{eqnarray}
g_i = \left( \frac{\partial E (\b{p})}{\partial p_i} \right)_{\b{p} = \b{p}^0},
\end{eqnarray}
are the components of the energy gradient vector $\b{g}$, and
\begin{eqnarray}
h_{ij} = \left( \frac{\partial^2 E (\b{p})}{\partial p_i \partial p_j} \right)_{\b{p} = \b{p}^0},
\end{eqnarray}
are the elements of the energy Hessian matrix $\b{h}$. Imposition of the
stationarity condition on the expanded energy expression, $\partial
E^{[2]} (\b{p}) /\partial p_i = 0$, gives the following
standard solution for the parameter variations
\begin{eqnarray}
 \Delta \b{p}= - \b{h}^{-1} \cdot \b{g},
\label{deltaalphanewton}
\end{eqnarray}
where $\b{h}^{-1}$ is the inverse of the Hessian matrix. In practice,
the energy gradient and Hessian are calculated in VMC with the
statistical estimators given in Sec.~\ref{sec:gradhess}, yielding
the parameter variations $\Delta \b{p}$ of
Eq.~(\ref{deltaalphanewton}) that are used to update the current wave
function, $\ket{\Psi_0} \to \ket{\Psi(\b{p}^0 + \Delta
  \b{p})}$. It simply remains to iterate until convergence.

\vskip 2mm \noindent {\it Stabilization.} \hskip 2mm
As explained in Ref.~\onlinecite{UmrFil-PRL-05}, the stabilization of
the Newton method is achieved by adding a positive constant,
$a_{\diag} \geq 0$, to the diagonal of the Hessian matrix $\b{h}$,
i.e., $h_{ij} \to h_{ij} + a_{\diag} \delta_{ij}$. As $a_{\diag}$ is
increased, the parameter variations $\Delta \b{p}$ become
smaller and rotate from the Newtonian direction to the steepest
descent direction. A good value of $a_{\diag}$ is
automatically determined at each iteration by performing three very
short Monte Carlo calculations using correlated sampling with wave
function parameters obtained with three trial values of $a_{\diag}$
and predicting by parabolic interpolation the value of $a_{\diag}$
that minimizes the energy~\cite{UmrTouFilSorHeg-JJJ-XX}, with some bounds imposed.
The use of correlated sampling makes it possible to calculate energy differences
with much smaller statistical error than the energies themselves.  This procedure
helps convergence if one is far from the minimum or if the
statistical noise is large in the Monte Carlo evaluation of the
gradient and Hessian.

We have found that adding in a multiple of the unit matrix to the Hessian as described
above works well, but there exist other possible choices of positive definite
matrices that could be added in. For instance, Sorella~\cite{Sor-PRB-05} adds in a multiple of the overlap matrix of the first-order derivatives of the wave function. Another possible choice is a multiple of the
Levenberg-Marquardt approximation to the Hessian of the variance of the local energy.

\subsection{Linear optimization method}
\label{sec:linear}

The most straightforward way to energy-optimize linear parameters in wave functions, such as the CSF parameters, is to diagonalize the Hamiltonian in the variational space that they define, leading to a generalized eigenvalue equation. This has been done in QMC for example in Refs.~\onlinecite{CepBer-JCP-88,NigMel-PRL-01}. The linear method that we present now is an extension of the approach of Ref.~\onlinecite{NigMel-PRL-01} to arbitrary nonlinear parameters. This method is also presented in Ref.~\onlinecite{UmrTouFilSorHeg-JJJ-XX}, using slightly different but equivalent conventions.

For notational convenience, we first introduce the normalized wave function
\begin{eqnarray}
\ket{\Psib(\b{p})} = \frac{\ket{\Psi(\b{p})}}{\sqrt{\braket{\Psi(\b{p})}{\Psi(\b{p})}}}.
\label{Psin}
\end{eqnarray}
The idea is then to expand this normalized wave function $\ket{\Psib(\b{p})}$ to first-order in the parameters $\b{p}$ around the current parameters $\b{p}^0$,
\begin{eqnarray}
\ket{\Psib_\lin(\b{p})} = \ket{\Psi_0} + \sum_{i=1}^{N^{\opt}} \Delta p_i \, \ket{\Psib_i},
\label{Psib1}
\end{eqnarray}
where the wave function at $\b{p}=\b{p}^0$ is simply $\ket{\Psib(\b{p}^0)} = \ket{\Psib_0} = \ket{\Psi_0}$ (chosen to be normalized to $1$) and, for $i \geq 1$, $\ket{\Psib_i}$ are the derivatives of $\ket{\Psib(\b{p})}$ that are orthogonal to $\ket{\Psi_0}$
\begin{eqnarray}
\ket{\Psib_i}= \left( \frac{\partial \ket{\Psib(\b{p})}}{\partial p_i} \right)_{\b{p}=\b{p}^0} = \ket{\Psi_i} - S_{0i} \, \ket{\Psi_0},
\end{eqnarray}
where $S_{0i}= \braket{\Psi_0}{\Psi_i}$. The minimization of the energy calculated with this linear wave function
\begin{eqnarray}
E_\lin = \min_{\b{p}} E_\lin (\b{p}),
\label{E1min}
\end{eqnarray}
where
\begin{eqnarray}
E_\lin (\b{p}) = \frac{\bra{\Psib_\lin(\b{p})} \hat{H} \ket{\Psib_\lin(\b{p})}}{\braket{\Psib_\lin(\b{p})}{\Psib_\lin(\b{p})}},
\label{Elin}
\end{eqnarray}
leads to the stationary condition of the associated Lagrange function
\begin{eqnarray}
\nabla_{\b{p}} \left[ \bra{\Psib_\lin(\b{p})} \hat{H} \ket{\Psib_\lin(\b{p})} - E_\lin \braket{\Psib_\lin(\b{p})}{\Psib_\lin(\b{p})} \right] = 0,
\nonumber\\
\label{nablaL}
\end{eqnarray}
where $E_\lin$ acts as a Lagrange multiplier for the normalization condition. The Lagrange function is quadratic in $\b{p}$ and Eq.~(\ref{nablaL}) leads to the following generalized eigenvalue equation
\begin{eqnarray}
\overline{\b{H}} \cdot \Delta \b{p} = E_\lin \, \overline{\b{S}} \cdot \Delta \b{p},
\label{GEQ}
\end{eqnarray}
where $\overline{\b{H}}$ is the matrix of the Hamiltonian $\hat{H}$ in the $(N^{\opt}+1)$-dimensional basis consisting of the current normalized wave function and its derivatives $\{\ket{\Psib_0}, \ket{\Psib_1},\ket{\Psib_2}, \cdots, \ket{\Psib_{N^{\opt}}} \}$, with elements $\overline{H}_{ij} = \bra{\Psib_i}\hat{H}\ket{\Psib_j}$, $\overline{\b{S}}$ is the overlap matrix of this $(N^{\opt}+1)$-dimensional basis, with elements $\overline{S}_{ij} = \braket{\Psib_i}{\Psib_j}$ (note that $\overline{S}_{00}=1$ and $\overline{S}_{i0}=\overline{S}_{0i}=0$ for $i \geq 1$), and $\Delta \b{p}$ is the $(N^{\opt}+1)$-dimensional vector of parameter variations with $\Delta p_0=1$.
The linear method consists of solving the generalized eigenvalue equation of Eq.~(\ref{GEQ}),
for the lowest (physically reasonable) eigenvalue and associated eigenvector denoted by $\Delta \overline{\b{p}}$.
The overlap and (non-symmetric) Hamiltonian matrices are computed in VMC using the statistical estimators
given in Sec.~\ref{sec:ovlpham}.
Although we focus here on the optimization of the ground-state wave function,
solving Eq.~(\ref{GEQ}) also gives upper bound estimates of excited state energies
of states with the same spatial and spin symmetries.

However, there is an arbitrariness in the previously described procedure: we have found
the parameter variations $\Delta \b{\pb}$ from the expansion of the wave function
$\ket{\Psib(\b{p})}$ of Eq.~(\ref{Psin}), but another choice of the normalization of the
wave function will lead to different parameter variations. To see that, consider
a differently-normalized wave function
\begin{equation}
\ket{\Psibb(\b{p})} = N(\b{p}) \ket{\Psib(\b{p})},
\label{Psibb}
\end{equation}
where the normalization function $N(\b{p})$ is chosen to satisfy $N(\b{p}^0)=1$ so as to leave unchanged the normalization at $\b{p}=\b{p}^0$, i.e. $\ket{\Psibb(\b{p}^0)} = \ket{\Psi_0}$. The derivatives of this new wave function are
\begin{eqnarray}
\ket{\Psibb_i}= \left( \frac{\partial \ket{\Psibb(\b{p})}}{\partial p_i} \right)_{\b{p}=\b{p}^0} = \ket{\Psib_i} + N_{i} \, \ket{\Psi_0},
\end{eqnarray}
where $N_{i} =\left(  \partial  N(\b{p}) / \partial p_i \right)_{\b{p}=\b{p}^0}$, i.e. their projections onto the current wave function $\ket{\Psi_0}$ depend on the normalization. Consequently, the first-order expansion of this new wave function
\begin{eqnarray}
\ket{\Psibb_\lin(\b{p})} = \ket{\Psi_0} + \sum_{i=1}^{N^{\opt}} \Delta p_i \, \ket{\Psibb_i},
\label{Psibb}
\end{eqnarray}
leads, after optimization of the energy, to different optimal parameter variations $\Delta \b{\pbb}$. As the two wave functions $\ket{\Psib_\lin(\b{p})}$ and $\ket{\Psibb_\lin(\b{p})}$ lie in the same variational space, they must be proportional after minimization of the energy, which implies that the new optimal parameter variations $\Delta \b{\pbb}$ are actually related to the original optimal parameter variations $\Delta \b{\pb}$ by a uniform rescaling
\begin{eqnarray}
\Delta \b{\pbb} = \frac{\Delta \b{\pb}}{1-\sum_{i=1}^{N^\opt} N_i \Delta \pb_i}.
\label{deltapbb}
\end{eqnarray}
Any choice of normalization does not necessarily give good parameter variations. For the CSF parameters, it is obvious that the best choice is the normalization of the wave function of Eq.~(\ref{Psi}) in order to keep the linear dependence on these parameters, ensuring convergence of the linear method in a single step. This is achieved by choosing $\ket{\Psibb_i}=\ket{\Psi_i}$ which gives
\begin{eqnarray}
N_i = S_{i0}, \,\,\,\, \text{for linear parameters}.
\label{Nil}
\end{eqnarray}
For the nonlinear Jastrow and orbital parameters, several criteria are possible. We have found that a good one is to choose the normalization by imposing that, for the variation of the nonlinear parameters, each derivative $\ket{\Psibb_i}$ is orthogonal to a linear combination of $\ket{\Psi_0}$ and $\ket{\Psib_\lin}$, i.e. $\braket{\Psibb_i}{\xi \Psi_0 + (1-\xi) \Psib_\lin/||\Psib_\lin||}=0$, where $\xi$ is a constant between $0$ and $1$, resulting in
\begin{eqnarray}
N_i &=& - \frac{(1-\xi) \sum_j^\text{nonlin} \Delta \pb_j \overline{S}_{ij}}{(1-\xi) +\xi \sqrt{1+\sum_{j,k}^\text{nonlin} \Delta \pb_j \Delta \pb_k \overline{S}_{jk}  }} ,
\nonumber\\
&&
 \text{for nonlinear parameters},
\label{Ninl}
\end{eqnarray}
where the sums are only over the nonlinear Jastrow and orbital parameters.
The simple choice $\xi=1$ first used by Sorella~\cite{Sor-PRB-01} in the context
of the SR method leads in many cases to good parameter variations, but in some cases
can result in parameter variations that are too large.
The choice $\xi=0$ making the norm of the linear wave function change
$\vert\vert \Psibb_{\lin} - \Psi_0 \vert\vert$ minimum is safer but in some cases
can yield parameter variations that are too small.
In those cases, the choice $\xi=1/2$, imposing $\vert\vert \Psibb_\lin \vert\vert = \vert\vert \Psi_0 \vert\vert$,
avoids both too large and too small parameter variations.
In particular, if $\Delta \b{\pb} = \infty$, meaning that $\Psib_\lin$ is orthogonal to $\Psi_0$, it follows from Eqs.~(\ref{deltapbb}) and (\ref{Ninl}) that $\Delta \b{\pbb}$ is zero for $\xi=0$ but $\Delta \b{\pbb}$ is nonzero and finite for $\xi=1/2$.
In practice, all these three choices for $\xi$ usually lead to a very rapid convergence of the nonlinear parameters. In contrast, choosing the original derivatives, i.e. $N_i = S_{i0}$, leads to slowly converging or diverging Jastrow parameters.

\vskip 2mm \noindent {\it Stabilization.} \hskip 2mm
Similarly to the procedure used for the Newton method, we stabilize
the linear method by adding a positive constant, $a_{\diag} \geq 0$,
to the diagonal of $\overline{\b{H}}$ except for the first element,
i.e. $\overline{H}_{ij} \to \overline{H}_{ij} + a_{\diag} \delta_{ij}
(1 -\delta_{i0})$. Again, as $a_{\diag}$ becomes larger, the parameter variations $\Delta \b{p}$ become
smaller and rotate toward the steepest descent direction.
The value of $a_{\diag}$ is then automatically
adjusted in the course of the optimization in the same way as in the
Newton method.
Note that if instead we were to add $a_{\diag}$ to $\overline{\b{S}}^{\, -1} \cdot \overline{\b{H}}$
then it would be the ``level-shift''
parameter commonly used in diagonalization procedures.
We prefer to add to $\overline{\b{H}}$, in part, because it is not necessary to
compute $\overline{\b{S}}^{\, -1} \cdot \overline{\b{H}}$ in order to solve
Eq.~(\ref{GEQ}).

\vskip 2mm \noindent {\it Connection to the EFP method.} \hskip 2mm
The generalized eigenvalue equation of Eq.~(\ref{GEQ}) can be re-written
as an eigenvalue equation,
$\overline{\b{H}}' \cdot \Delta \b{p} = E_\lin \Delta \b{p}$,
where $\overline{\b{H}}' = \overline{\b{S}}^{\, -1} \cdot \overline{\b{H}}$,
i.e. with matrix elements
$\overline{H}_{ij}' = \sum_{k=0}^{N^\opt} ( \overline{\b{S}}^{\, -1} )_{ik} \bra{\Psib_k} \hat{H} \ket{\Psib_j}$.
This form is useful to establish the connection with the EFP optimization method
for the CSF and orbital parameters~\cite{FilFah-JCP-00,SchFah-JCP-02,SchFil-JCP-04}.
This latter approach consists of solving at each iteration the
effective eigenvalue equation,
$\overline{\b{H}}^{\EFP} \cdot \Delta \b{p} = E^{\EFP} \Delta \b{p}$,
where the EFP effective Hamiltonian has matrix elements
$\overline{H}^{\EFP}_{ij} = \bra{\Phi_i} \hat{H} \ket{\Phi_i} \delta_{ij} + \sum_{k=1}^{N^\opt} ( \overline{\b{S}}^{\, -1} )_{ik} \bra{\Psib_k} \hat{H} \ket{\Psib_0}  \left[ (1-\delta_{i0}) \delta_{0j}  + \delta_{i0} (1-\delta_{0j}) \right]$,
where $\ket{\Phi_i}$ designates the current wave function and its derivatives
without the Jastrow factor, i.e. $\ket{\Psi_i} = \hat{J}(\bm{\alpha}^0) \ket{\Phi_i}$, and $\bra{\Psib_k} \hat{H} \ket{\Psib_0}$ are just the components of half the gradient of the energy. Hence, in the EFP method, only the off-diagonal elements in the first
column and first row calculated from the components of the energy gradient are retained in $\overline{\b{H}}^{\EFP}$.

\vskip 2mm \noindent {\it Connection to the Newton and SRH methods.} \hskip 2mm
In the linear method, the energy expression that is minimized at each iteration,
$E_\lin(\b{p})$, contains all orders in the parameter variations because of the
presence of the denominator in Eq.~(\ref{Elin}), though only the zeroth- and first-order terms
match those of the expansion of the exact energy $E(\b{p})$. In contrast, in the Newton method,
the energy expression of Eq.~(\ref{E2}), $E^{[2]}(\b{p})$, is truncated at second order in
$\Delta \b{p}$ but is exact up to this order. Now, if instead of solving the generalized eigenvalue
equation~(\ref{GEQ}), one expands the energy expression of Eq.~(\ref{Elin}) to second order in
$\Delta \b{p}$, one recovers the Newton method with an approximate (symmetric) Hessian
$h_{ij}^{\lin}=\overline{H}_{ij}+\overline{H}_{ji} - 2 E_0 \overline{S}_{ij}$ corresponding
exactly to the SRH method with $\beta=0$ of Ref.~\onlinecite{Sor-PRB-05}.
The SRH method is much less stable and converges more slowly than either our linear method
or our Newton method for the systems studied here.

\subsection{Perturbative optimization method}

The perturbative method discussed next is identical to the perturbative
EFP approach of Scemama and Filippi~\cite{SceFil-PRB-06} for the
optimization of the CSF and orbital parameters, provided that the same
choice is made for the energy denominators (see below).
We give here an alternate proof without introducing the concept of energy
fluctuations that in principle extends the method to other kinds of
parameters as well.

Instead of calculating the optimal linearized wave function $\ket{\Psib_\lin}$
by diagonalizing the Hamiltonian $\hat{H}$ in the subspace spanned by
$\{\ket{\Psib_0}, \ket{\Psib_1}, \ket{\Psib_2}, \cdots, \ket{\Psib_{N^{\opt}}} \}$,
we formulate a nonorthogonal perturbation theory for $\ket{\Psib_\lin}$.
The textbook formulation of perturbation theory starts from the
Hamiltonian $\Hhat$ whose eigenstates we wish to compute, and a zeroth order
Hamiltonian $\Hhat^{(0)}$ whose eigenstates are known.
Instead, here we start with $\Hhat$ and the states
$\{\ket{\Psib_0}, \ket{\Psib_1}, \ket{\Psib_2}, \cdots, \ket{\Psib_{N^{\opt}}} \}$,
and define a zeroth order operator $\Hhat^{(0)}$ for which these states
are right eigenstates.
To do this, we introduce $\{\bra{\tilde{\Psi}_i}\}$, the dual
(biorthonormal) basis of the basis $\{\ket{\Psib_i}\}$, i.e.
$\braket{\tilde{\Psi}_i}{\Psib_j}=\delta_{ij}$, given by (see, e.g.,
Ref.~\onlinecite{ArtMil-PRA-91})
\begin{equation}
\bra{\tilde{\Psi}_i} = \sum_{j=0}^{N^{\opt}} (\overline{\b{S}}^{\, -1})_{ij} \, \bra{\Psib_j},
\end{equation}
where $(\overline{\b{S}}^{\, -1})_{ij}$ are the elements of the
inverse of the overlap matrix $\overline{\b{S}}$, and we introduce the
non-Hermitian projector operator onto this subspace
\begin{equation}
\hat{P} = \sum_{i=0}^{N^{\opt}} \ket{\Psib_i} \bra{\Psit_i}.
\end{equation}
The optimal linearized wave function, minimizing the energy
[Eq.~(\ref{E1min})], satisfies the projected Schr\"odinger equation
\begin{equation}
\hat{P} \hat{H} \ket{\Psib_\lin} = E_\lin \hat{P} \ket{\Psib_\lin},
\label{PHPsiEPPsi}
\end{equation}
with the normalization condition $\braket{\Psit_0}{\Psib_\lin}=1$,
ensuring that the coefficient of $\ket{\Psib_\lin}$ on $\ket{\Psib_0}=\ket{\Psi_0}$
is $1$ as in Eq.~(\ref{Psib1}).

To construct the perturbation theory, we now introduce a fictitious
projected Schr\"odinger equation depending on a coupling constant $\l$
\begin{equation}
\hat{P} \hat{H}^{\l} \ket{\Psib_\lin^\l} = E_\lin^\l \hat{P} \ket{\Psib_\lin^\l},
\label{PHlPsilElPPsil}
\end{equation}
with the normalization condition $\braket{\Psit_0}{\Psib_\lin^\l}=1$
for all $\l$, so that, for $\l=1$, Eq.~(\ref{PHlPsilElPPsil}) reduces
to Eq.~(\ref{PHPsiEPPsi}): $\hat{H}^{\l=1} = \hat{H}$,
$\ket{\Psib_\lin^{\l=1}}=\ket{\Psib_\lin}$, $E_\lin^{\l=1}=E_\lin$,
and we partition the Hamiltonian $\hat{H}^{\l}$ as follows
\begin{equation}
\hat{H}^{\l} = \hat{H}^{(0)} + \l \hat{H}^{(1)}.
\end{equation}
In this expression, $\hat{H}^{(0)}$ is a zeroth-order non-Hermitian
operator
\begin{equation}
\hat{H}^{(0)} = \sum_{i=0}^{N^{\opt}} {\cal E}_i  \ket{\Psib_i} \bra{\tilde{\Psi}_i},
\end{equation}
where ${\cal E}_i$ are \textit{arbitrary} energies. Clearly,
$\hat{H}^{(0)}$ admits $\ket{\Psib_i}$ as right-eigenstate and
$\bra{\tilde{\Psi}_i}$ as left-eigenstate, with common eigenvalue
${\cal E}_i$. The non-Hermitian perturbation operator is obviously
defined as $\hat{H}^{(1)}=\hat{H}-\hat{H}^{(0)}$. We expand
$\ket{\Psib_\lin^\l}$ and $E_\lin^\l$ in powers of $\l$:
$\ket{\Psib_\lin^\l} =\sum_{k=0}^{\infty} \l^k \ket{\Psib_\lin^{(k)}}$
and $E_\lin^\l =\sum_{k=0}^{\infty} \l^k E_\lin^{(k)}$. The
zeroth-order (right) eigenstate and energy are simply:
$\ket{\Psib_\lin^{(0)}} = \ket{\Psib_0}$ and $E_\lin^{(0)}= {\cal E}_0$.
The first-order correction to the wave function is determined
by the equation
\begin{equation}
\hat{P} \left( \hat{H}^{(0)} - {\cal E}_0 \right) \ket{\Psib_\lin^{(1)}} = - \hat{P} \left(  \hat{H}^{(1)} - E_\lin^{(1)}  \right) \ket{\Psib_0}.
\label{Psi1eq}
\end{equation}
To solve this equation, we define the non-Hermitian projector operator
$\hat{R}=\sum_{i=1}^{N^\opt} \ket{\Psib_i}\bra{\Psit_i}$ which, in
comparison to the projector $\hat{P}$, also removes the component parallel to
$\ket{\Psib_0}$. Note that $\hat{R}\hat{P}=\hat{R}$, $\hat{R}$
commutes with $\hat{H}^{(0)} - {\cal E}_0$ and $\hat{R}
\ket{\Psib_\lin^{(1)}} = \ket{\Psib_\lin^{(1)}}$ (since
$\braket{\Psit_0}{\Psib_\lin^{\l}}=1$ and
$\braket{\Psit_0}{\Psib_0}=1$, implying
$\braket{\Psit_0}{\Psib_\lin^{(1)}}=0$), so that applying $\hat{R}$ on
Eq.~(\ref{Psi1eq}) leads to
\begin{eqnarray}
 \ket{\Psib_\lin^{(1)}} &=& - \frac{\hat{R}}{\hat{H}^{(0)} - {\cal E}_0 } \left(  \hat{H}^{(1)} - E_\lin^{(1)} \right) \ket{\Psib_0}
\nonumber\\
&=& - \sum_{i=1}^{N^{\opt}} \ket{\Psib_i} \frac{\bra{\Psit_i} \hat{H} - {\cal E}_0 - E_\lin^{(1)} \ket{\Psib_0}}{{\cal E}_i - {\cal E}_0}
\nonumber\\
&=&
- \sum_{i=1}^{N^{\opt}} \sum_{j=1}^{N^{\opt}}  (\overline{\b{S}}^{\, -1})_{ij} \frac{\bra{\Psib_j} \hat{H} \ket{\Psib_0}}{{\cal E}_i - {\cal E}_0} \ket{\Psib_i},
\label{}
\end{eqnarray}
where ${\cal E}_0$ and $E_\lin^{(1)}$ in the numerator and the term
$j=0$ have been dropped since, for $i \neq 0$,
$\braket{\Psit_i}{\Psib_0}=0$ and $(\overline{\b{S}}^{\, -1})_{i0}
=0$, respectively. Therefore, the parameter variations in this
first-order perturbation theory are
\begin{equation}
\Delta \pb_i^{(1)} = - \frac{1}{\Delta {\cal E}_i} \sum_{j=1}^{N^{\opt}} (\overline{\b{S}}^{\, -1})_{ij} \overline{H}_{j0},
\label{Deltaalpha1}
\end{equation}
where $\overline{H}_{j0}=\bra{\Psib_j} \hat{H} \ket{\Psib_0} = \bra{\Psi_j} \hat{H} - E_0\ket{\Psi_0} = g_j/2$ is just half the gradient of the energy and $\Delta {\cal E}_i = {\cal E}_i - {\cal E}_0$. The perturbative method consists of calculating the parameter variations $\Delta \b{\pb}^{(1)}$ according to Eq.~(\ref{Deltaalpha1}), updating the current wave function, $\ket{\Psi_0} \to \ket{\Psi(\b{p}^0 + \Delta \b{\pb}^{(1)})}$, and iterating until convergence.
It is apparent from Eq.~(\ref{Deltaalpha1}) that the perturbative method can be viewed
as the Newton method with an approximate Hessian,
$h_{ij}^{\pert} = (\overline{\b{S}}^{\, -1})_{ij}/\Delta {\cal E}_i$,
as also noted in Ref.~\onlinecite{SceFil-PRB-06}.

The energy denominators $\Delta {\cal E}_i$ in Eq.~(\ref{Deltaalpha1})
remain to be chosen.
Since perturbation theory works best when $\hat{H}^{(0)}$ is ``close'' to $\hat{H}$
we choose $\hat{H}^{(0)}$ to have the same diagonal elements as $\hat{H}$, resulting in
\begin{equation}
\Delta {\cal E}_i = \frac{\bra{\Psib_i} \hat{H} \ket{\Psib_i}}{\braket{\Psib_i}{\Psib_i}} -E_0 = \frac{\overline{H}_{ii}}{\overline{S}_{ii}} - \overline{H}_{00}.
\label{DeltaEi}
\end{equation}
In practice, only rough estimates of the $\Delta {\cal E}_i$'s are necessary for the optimization so that one can compute them for just
the initial iteration and keep them fixed for the following
iterations. Therefore, for these iterations, only the inverse overlap
matrix, $\overline{\b{S}}^{\, -1}$, and the gradient of the energy,
$g_j = 2 \, \overline{H}_{j0}$, need to be calculated in the
perturbative method, leading to an important computational speedup
per iteration in comparison to the linear method.

\vskip 2mm \noindent {\it Stabilization.} \hskip 2mm
Similarly to the linear method, the perturbative method can be
stabilized by adding an adjustable positive constant, $a_{\diag} \geq
0$, to the energy denominators, i.e. $\Delta {\cal E}_i \to \Delta
{\cal E}_i + a_{\diag}$, which has the effect of decreasing the
parameter variations $\Delta \b{\pb}^{(1)}$.

\vskip 2mm \noindent {\it Connection to the perturbative EFP and SR methods.} \hskip 2mm
For the CSF and orbital parameters, if the energy
denominators are chosen to be $\Delta {\cal E}_{i} = \bra{\Phi_i} \hat{H}
\ket{\Phi_i} / \braket{\Phi_i}{\Phi_i} - \bra{\Phi_0} \hat{H}
\ket{\Phi_0} / \braket{\Phi_0}{\Phi_0}$ (i.e., without the Jastrow
factor), Eq.~(\ref{Deltaalpha1}) exactly reduces to the perturbative
EFP method~\cite{SceFil-PRB-06}. Also, Eq.~(\ref{Deltaalpha1}) reduces
to the SR optimization method~\cite{Sor-PRB-01,CasSor-JCP-03,CasAttSor-JCP-04} if the energy
denominators are all chosen equal, $\Delta {\cal E}_i = \Delta {\cal E}$ for all $i$.

\section{Variational Monte Carlo realization}
\label{sec:vmc}

When the previously-described energy minimization procedures are implemented
in VMC it is important to pay attention to the statistical fluctuations.
Expressions that are equivalent in the limit of an infinite Monte Carlo
sample, can in fact have very different statistical errors for a finite sample.
We provide prescriptions for low variance estimators in this section.

We also note that, in order to reduce round-off noise, it can help to rescale the
elements of the gradient vector, and the hessian, Hamiltonian and overlap matrices
using the square root of the diagonal of overlap matrix.

At each step of the optimization, the quantum-mechanical
averages are computed by sampling the probability density of the
current wave function $\Psi_0(\b{R})^2$. We will denote the
statistical average of a local quantity, $f(\b{R})$, by $\left\langle
  f(\b{R}) \right\rangle = (1/M) \sum_{k=1}^{M} f(\b{R}_k)$ where the
$M$ electron configurations $\b{R}_k$ are sampled from
$\Psi_0(\b{R})^2$.

\subsection{Energy gradient and Hessian}
\label{sec:gradhess}

In terms of the derivatives $\Psi_i(\b{R})$ of the wave function of Eq.~(\ref{Psi}), and using the Hermiticity of the Hamiltonian $\hat{H}$, an estimator of the energy gradient is ~\cite{CepCheKal-PRB-77}
\begin{equation}
g_i = 2 \Biggl[ \left\langle \frac{\Psi_i(\b{R})}{\Psi_0(\b{R})} E_\L(\b{R}) \right\rangle - \left\langle \frac{\Psi_i(\b{R})}{\Psi_0(\b{R})} \right\rangle \left\langle E_\L(\b{R}) \right\rangle \Biggl],
\label{gi}
\end{equation}
where $E_\L(\b{R})=\left[ H(\b{R})\Psi_0(\b{R}) \right]/\Psi_0(\b{R})$ is the local energy. In the limit that $\Psi_0(\b{R})$ is an exact eigenfunction, the local energy becomes constant, $E_\L(\b{R})= E_\exact$ for all $\b{R}$, and thus the gradient of Eq.~(\ref{gi}) vanishes with zero variance. This leads to the following zero-variance principle for the Newton and perturbative methods: in the limit that $\Psi_0(\b{R})$ is an exact eigenfunction, the parameter variations of Eqs.~(\ref{deltaalphanewton}) and~(\ref{Deltaalpha1}) vanish with zero variance.

Taking the derivative of Eq.~(\ref{gi}) leads to the straightforward estimator of the energy Hessian of Lin, Zhang and Rappe (LZR)~\cite{LinZhaRap-JCP-00}
\begin{equation}
h_{ij}^{\text{LZR}} = A_{ij} + B_{ij} + C_{ij},
\label{hijLZR}
\end{equation}
where
\begin{eqnarray}
A_{ij} &=& 2 \Biggl[  \left\langle \frac{\Psi_{ij}(\b{R})}{\Psi_0(\b{R})} E_\L(\b{R}) \right\rangle - \left\langle \frac{\Psi_{ij}(\b{R})}{\Psi_0(\b{R})} \right\rangle \left\langle E_\L(\b{R}) \right\rangle
\nonumber\\
&& - \left\langle \frac{\Psi_i(\b{R})}{\Psi_0(\b{R})} \frac{\Psi_j(\b{R})}{\Psi_0(\b{R})}  E_\L(\b{R}) \right\rangle
\nonumber\\
&& + \left\langle  \frac{\Psi_i(\b{R})}{\Psi_0(\b{R})} \frac{\Psi_j(\b{R})}{\Psi_0(\b{R})} \right\rangle \left\langle E_\L(\b{R}) \right\rangle \Biggl],
\nonumber\\
\label{Aij}
\end{eqnarray}
involving the second derivatives $\Psi_{ij}(\b{R})$ of the wave function,
\begin{eqnarray}
B_{ij} &=&  4 \Biggl[\left\langle \frac{\Psi_i(\b{R})}{\Psi_0(\b{R})} \frac{\Psi_j(\b{R})}{\Psi_0(\b{R})}  E_\L(\b{R}) \right\rangle
\nonumber\\
&& - \left\langle \frac{\Psi_i(\b{R})}{\Psi_0(\b{R})} \frac{\Psi_j(\b{R})}{\Psi_0(\b{R})} \right\rangle \left\langle E_\L(\b{R}) \right\rangle \Biggl]
\nonumber\\
&&-  2 \left\langle \frac{\Psi_{i}(\b{R})}{\Psi_0(\b{R})} \right\rangle g_j
-  2 \left\langle \frac{\Psi_{j}(\b{R})}{\Psi_0(\b{R})} \right\rangle g_i
\nonumber\\
&=& 4 \, \Bigg\langle \left( \frac{\Psi_i(\b{R})}{\Psi_0(\b{R})} - \left\langle \frac{\Psi_i(\b{R})}{\Psi_0(\b{R})} \right\rangle \right)
 \nonumber \\
 &&  \times  \left( \frac{\Psi_j(\b{R})}{\Psi_0(\b{R})} - \left\langle \frac{\Psi_j(\b{R})}{\Psi_0(\b{R})} \right\rangle \right)
\left( E_\L(\b{R}) - \left\langle E_\L(\b{R}) \right\rangle \right) \Bigg\rangle,
 \nonumber \\
\label{Bij}
\end{eqnarray}
and
\begin{eqnarray}
C_{ij} &=&  2 \left\langle \frac{\Psi_{i}(\b{R})}{\Psi_0(\b{R})} E_{\L,j}(\b{R}) \right\rangle,
\label{Cij}
\end{eqnarray}
where $E_{\L,j}(\b{R}) = \left[ H(\b{R})\Psi_j(\b{R})
\right]/\Psi_0(\b{R}) - \left[ \Psi_j(\b{R})/\Psi_0(\b{R}) \right]
E_\L(\b{R})$ is the derivative of the local energy with respect to parameter $j$. In this estimator
of the Hessian, the term that fluctuates the most is $C_{ij}$.

Umrigar and Filippi~\cite{UmrFil-PRL-05} observed that the fluctuations of
a covariance $\langle ab \rangle - \langle a \rangle \langle b \rangle$ are much
smaller than those of $\langle ab \rangle$ if
the fluctuations of $a$ are much smaller than the average of $a$, i.e.,
$\sqrt{\langle a^2 \rangle - \langle a \rangle ^2} \ll \vert \langle a \rangle \vert$,
and, $a$ is not strongly correlated with $1/b$.
In Eq.~(\ref{Cij}), $\Psi_{i}(\b{R})/\Psi_0(\b{R})$ is always of the same sign
for parameters in the exponent and in practice its fluctuations are much smaller than its average.
Further, it follows from the Hermiticity of the Hamiltonian that $\left\langle E_{\L,j}(\b{R}) \right\rangle$
vanishes in the limit of an infinite sample~\cite{LinZhaRap-JCP-00}.
Using these two observations, Umrigar and Filippi~\cite{UmrFil-PRL-05} provided an
estimator of the Hessian,
\begin{equation}
h_{ij}^{\text{UF}} = A_{ij} + B_{ij} + D_{ij},
\label{hijUF}
\end{equation}
that fluctuates much less than the straightforward LZR estimator, where the symmetrized estimator
\begin{eqnarray}
D_{ij} &=&   \left\langle \frac{\Psi_{i}(\b{R})}{\Psi_0(\b{R})} E_{\L,j}(\b{R}) \right\rangle - \left\langle \frac{\Psi_{i}(\b{R})}{\Psi_0(\b{R})} \right\rangle \left\langle E_{\L,j}(\b{R}) \right\rangle
\nonumber\\
&&+  \left\langle \frac{\Psi_{j}(\b{R})}{\Psi_0(\b{R})} E_{\L,i}(\b{R}) \right\rangle - \left\langle \frac{\Psi_{j}(\b{R})}{\Psi_0(\b{R})} \right\rangle \left\langle E_{\L,i}(\b{R}) \right\rangle,
\nonumber\\
\label{Dij}
\end{eqnarray}
has the same average as the term $C_{ij}$ in the limit of an infinite sample, but being a covariance has much smaller fluctuations.
We note that $A_{ij}$ is already a covariance and $B_{ij}$ is a tri-covariance.

Although the $A_{ij}$ and $B_{ij}$ terms vanish with zero variance in the limit that $\Psi_0(\b{R})$
is an exact eigenfunction (the $D_{ij}$ term does not), in practice for the Jastrow parameters, far from the minimum,
the $B_{ij}$ fluctuates more than the $D_{ij}$ term for the Jastrow parameters in the Hessian of Eq.~(\ref{hijUF}).
With the form of the Jastrow factors that we use, we have observed that the ratio
$(B_{ij} + D_{ij})/D_{ij}$ is roughly independent of $i$ and $j$ for most $i$ and $j$ though it changes
during the Monte Carlo iterations.
It is typically between 1.2 and 2.5 at the initial iteration and between 0.9 and 1.1 at the final iteration.
We exploit this to decrease the fluctuations
by defining a new, approximate Hessian partially averaged over the Jastrow parameters
\begin{equation}
h_{ij}^{\text{TU}} = A_{ij} + \frac{\left\langle \left\langle |B_{ij}  + D_{ij}|\right\rangle \right\rangle}{\left\langle \left\langle |D_{ij}|\right\rangle \right\rangle} D_{ij},
\label{hijTU}
\end{equation}
where TU are the initials of the present authors, and the average over the Jastrow parameter pairs are defined by $\left\langle \left\langle X_{ij} \right\rangle \right\rangle = (2/N^\opt_\Jas (N^\opt_\Jas +1)) \sum_{i=1}^{N^\opt_\Jas} \sum_{j=i}^{N^\opt_\Jas} X_{ij}$. The average is
calculated as $\left\langle \left\langle |B_{ij} + D_{ij}| \right\rangle \right\rangle/\left\langle \left\langle |D_{ij}| \right\rangle
\right\rangle$ and not as $\left\langle \left\langle (|B_{ij} + D_{ij}|)/|D_{ij}| \right\rangle \right\rangle$ to avoid possible
numerical divergences of this ratio for small $D_{ij}$. In Eq.~(\ref{hijTU}), $i$ and $j$ refer only to Jastrow parameters. For
all the terms related to the other parameters (including all the mixed terms), the Hessian of Eq.~(\ref{hijUF}) is used without further
modification.

Exact or approximate wave functions such as $\Psi_0(\b{R})$ go linearly to zero with the distance $d$
between $\b{R}$ and their nodal hypersurface, i.e., $\Psi_0(\b{R}) \sim d$ for $d\to0$. The local energy
 $E_\L(\b{R})$ generally diverges as $1/d$ for $d\to0$ for approximate wave functions. In contrast to the case
of the Jastrow parameters, the derivatives $\Psi_i(\b{R})$ for the CSF and orbital parameters
have a different nodal hypersurface than $\Psi_0(\b{R})$ and the
ratio $\Psi_i(\b{R})/\Psi_0(\b{R})$ thus also diverges as $1/d$, even if the wave function $\Psi_0(\b{R})$ is exact.
Consequently, the derivative of the local energy $E_{\L,i}(\b{R})$ generally diverges as $1/d^2$ for approximate wave functions.
In the expression of the Hessian, the leading divergence at the nodes of the approximate wave function $\Psi_0(\b{R})$ thus comes from the terms
$\left(\Psi_i(\b{R})/\Psi_0(\b{R})\right) \left(\Psi_j(\b{R})/\Psi_0(\b{R})\right) E_\L(\b{R})$, $\left(\Psi_i(\b{R})/\Psi_0(\b{R})\right) E_{\L,j}(\b{R})$
and $\left(\Psi_j(\b{R})/\Psi_0(\b{R})\right) E_{\L,i}(\b{R})$ that behave as $1/d^3$.
It is however easy to check that these third-order divergences cancel exactly in Eq.~(\ref{hijUF}).

\subsection{Overlap and Hamiltonian matrices}
\label{sec:ovlpham}

The elements of the symmetric overlap matrix $\overline{\b{S}}$ are
\begin{subequations}
\begin{equation}
\overline{S}_{00}=1,
\label{}
\end{equation}
and, for $i, j > 0$,
\begin{equation}
\overline{S}_{i0}=\overline{S}_{0j}=0,
\label{}
\end{equation}
and
\begin{equation}
\overline{S}_{ij}=\left\langle \frac{\Psi_i(\b{R})}{\Psi_0(\b{R})} \frac{\Psi_j(\b{R})}{\Psi_0(\b{R})} \right\rangle  - \left\langle \frac{\Psi_i(\b{R})}{\Psi_0(\b{R})} \right\rangle \left\langle \frac{\Psi_j(\b{R})}{\Psi_0(\b{R})} \right\rangle.
\label{Sij}
\end{equation}
\label{S}
\end{subequations}

The elements of the Hamiltonian matrix $\overline{\b{H}}$ are
\begin{subequations}
\begin{equation}
\overline{H}_{00}= \left\langle E_\L(\b{R}) \right\rangle,
\label{}
\end{equation}
and, for $i, j > 0$,
\begin{equation}
\overline{H}_{i0}= \left\langle \frac{\Psi_i(\b{R})}{\Psi_0(\b{R})} E_\L(\b{R}) \right\rangle - \left\langle \frac{\Psi_i(\b{R})}{\Psi_0(\b{R})} \right\rangle \left\langle E_\L(\b{R}) \right\rangle,
\label{Hi0}
\end{equation}
\begin{eqnarray}
\overline{H}_{0j} &=& \left[ \left\langle \frac{\Psi_j(\b{R})}{\Psi_0(\b{R})} E_\L(\b{R}) \right\rangle - \left\langle \frac{\Psi_j(\b{R})}{\Psi_0(\b{R})} \right\rangle \left\langle E_\L(\b{R}) \right\rangle \right]
\nonumber\\
&&+ \left\langle E_{\L,j}(\b{R}) \right\rangle,
\label{H0j}
\end{eqnarray}
which are two estimators of half of the energy gradient, and
\begin{eqnarray}
\overline{H}_{ij} &=& \Biggl[ \left\langle \frac{\Psi_i(\b{R})}{\Psi_0(\b{R})} \frac{\Psi_j(\b{R})}{\Psi_0(\b{R})} E_\L(\b{R}) \right\rangle
\nonumber\\
&& - \left\langle \frac{\Psi_i(\b{R})}{\Psi_0(\b{R})} \right\rangle  \left\langle \frac{\Psi_j(\b{R})}{\Psi_0(\b{R})} E_\L(\b{R}) \right\rangle
\nonumber\\
&& - \left\langle \frac{\Psi_j(\b{R})}{\Psi_0(\b{R})} \right\rangle  \left\langle \frac{\Psi_i(\b{R})}{\Psi_0(\b{R})} E_\L(\b{R}) \right\rangle
\nonumber\\
&& + \left\langle \frac{\Psi_i(\b{R})}{\Psi_0(\b{R})} \right\rangle  \left\langle \frac{\Psi_j(\b{R})}{\Psi_0(\b{R})} \right\rangle  \left\langle E_\L(\b{R}) \right\rangle \Biggl]
\nonumber\\
&&+ \Biggl[ \left\langle \frac{\Psi_i(\b{R})}{\Psi_0(\b{R})} E_{\L,j}(\b{R}) \right\rangle - \left\langle \frac{\Psi_i(\b{R})}{\Psi_0(\b{R})} \right\rangle \left\langle E_{\L,j}(\b{R}) \right\rangle  \Biggl].
\nonumber\\
\label{Hij}
\end{eqnarray}
\label{H}
\end{subequations}
We do not use the Hermiticity of the Hamiltonian $\hat{H}$ to symmetrize the matrix $\overline{\b{H}}$. In fact, as shown by Nightingale and Melik-Alaverdian~\cite{NigMel-PRL-01}, using the non-symmetric matrix $\overline{\b{H}}$ of Eqs.~(\ref{H}) leads to a stronger zero-variance principle than the one previously-described for the Newton and perturbative methods: in the limit that the states $\{\ket{\Psib_0}, \ket{\Psib_1}, \ket{\Psib_2}, \cdots, \ket{\Psib_{N^{\opt}}} \}$ span an invariant subspace of the Hamiltonian $\hat{H}$, i.e. in the limit that the linear wave function $\Psib_\lin(\b{R})$ of Eq.~(\ref{Psib1}) after optimization is an exact eigenfunction, the matrix $\overline{\b{S}}^{\, -1} \cdot \overline{\b{H}}$ and consequently the eigenvector solution $\Delta \b{\pb}$ have zero variance. In practice, even if we do not work in an invariant subspace of $\hat{H}$, using the non-symmetric matrix $\overline{\b{H}}$ leads to smaller statistical errors on a finite sample than using its symmetrized analog. Although in principle diagonalization of a non-symmetric matrix leads to complex eigenvalues, in practice the physically reasonable (i.e., with large overlap with the current wave function) lowest eigenvectors have usually real eigenvalues. Of course, in the limit of an infinite sample $M \to \infty$ a symmetric matrix $\overline{\b{H}}$ is recovered.

As noted in the previous subsection for the Hessian, although the terms $\left(\Psi_i(\b{R})/\Psi_0(\b{R})\right) \left(\Psi_j(\b{R})/\Psi_0(\b{R})\right) E_\L(\b{R})$ and $\left(\Psi_i(\b{R})/\Psi_0(\b{R})\right) E_{\L,j}(\b{R})$ in the expression of the Hamiltonian matrix of Eq.~(\ref{Hij}) display a third-order divergence $1/d^3$ as the distance $d$ between $\b{R}$ and the nodal hypersurface of $\Psi_0(\b{R})$ goes to zero, again these divergences cancel exactly.

\subsection{Comparison of computational cost per iteration}

At each optimization iteration, besides the calculation the current wave function $\Psi_0(\b{R})$ and the local energy $E_{\L}(\b{R})$, the Newton method requires the computation of the first-order and second-order wave function derivatives, $\Psi_i(\b{R})$ and $\Psi_{ij}(\b{R})$, and the first-order derivatives of the local energy $E_{\L,i}(\b{R})$. The linear method requires the calculation of $\Psi_i(\b{R})$ and $E_{\L,i}(\b{R})$ but not of the second-order derivatives of the wave function with respect to the parameters. In principle, this decreases the computational cost per iteration, especially if the many orbital-orbital second-order derivatives were to be computed in the Newton method. In practice, since our implementation of the Newton method neglects these orbital-orbital derivatives, the computational cost per iteration of the Newton and linear methods is very similar.

The perturbative method requires the computation of the same quantities as the linear method.
However, since the method is not very sensitive to having accurate energy denominators
$\Delta {\cal E}_i$ in Eq.~(\ref{Deltaalpha1}),
and since the energy denominators do not undergo large changes from iteration to iteration
we compute these for the first iteration only.
Hence it is not necessary to compute $E_{\L,i}(\b{R})$ for subsequent iterations.
This leads to a computational speedup per iteration in comparison with the linear method.
The precise speedup factor depends on the wave function used; typically, for the systems studied here,
we have found factors ranging from about $1.5$ for a single-determinant wave function to
$5.5$ for the largest multi-determinant wave function considered, for the iterations
for which the $\Delta {\cal E}_i$'s are not computed.

\section{Computational details}
\label{sec:compdetails}

We illustrate the optimization methods by calculating the
ground-state electronic energy of the all-electron C$_2$ molecule at
the experimental equilibrium interatomic distance of $2.3481$
Bohr~\cite{CadWah-ADNDT-74}. The ground-state wave function is of
symmetry $^1\Sigma_g^+$ in the point group $D_{\infty h}$.
The estimated exact, infinite nuclear mass,
nonrelativistic electronic energy is $-75.9265(8)$
Hartree~\cite{BytRue-JCP-05}, where the number in parentheses is an
estimate of the uncertainty in the last digit. This system has a
strong multiconfiguration character due to the energetic near-degeneracy
of the valence orbitals, making it a challenging
system despite its small size.

We start by generating a standard {\it ab initio} wave function using the quantum chemistry program GAMESS~\cite{SchBalBoaElbGorJenKosMatNguSuWinDupMon-JCC-93}, typically a restricted Hartree-Fock (RHF) wave function or a MCSCF wave function, using the symmetry point group $D_{4h}$ which is the largest subgroup of $D_{\infty h}$ available in GAMESS.i
We use the uncontracted Slater basis set form of Clementi and Roetti~\cite{CleRoe-ADNDT-74}, with exponents reoptimized at the RHF level by Koga {\it et al.}~\cite{KogTatTha-PRA-93}. For carbon, the basis set contains two $1s$, three $2s$, one $3s$ and four $2p$ Slater functions, that are each approximated by a fit to six Gaussian functions~\cite{HehStePop-JCP-69,Ste-JCP-70} in GAMESS. Specifically, we consider the following {\it ab initio} wave functions: a RHF wave function, with orbital occupations $1\sigma_g^2 1\sigma_u^2 2\sigma_g^2 2\sigma_u^2 1\pi_{u,x}^2 1\pi_{u,y}^2$; a CAS(8,5) wave function, containing 6 CSFs in $D_{4h}$ symmetry made of 7 Slater determinants generated by distributing the 8 valence electrons over the 5 active valence orbitals $2\sigma_g 2\sigma_u 1\pi_{u,x} 1\pi_{u,y} 3\sigma_g$; a CAS(8,7) wave function, containing 80 CSFs made of 165 determinants with the 7 active orbitals $2\sigma_g 2\sigma_u 1\pi_{u,x} 1\pi_{u,y} 3\sigma_g 1\pi_{g,x} 1\pi_{g,y}$; a CAS(8,8) wave function, containing 264 CSFs made of 660 determinants with the 8 active orbitals $2\sigma_g 2\sigma_u 1\pi_{u,x} 1\pi_{u,y} 3\sigma_g 1\pi_{g,x} 1\pi_{g,y} 3\sigma_u$, i.e. all the valence orbitals originating from the $n=2$ shell of the C atoms. In addition, we construct a larger one-electron basis set by adding to the basis of Koga {\it et al.} one $d$ function with an exponent of $2.13$ optimized in RHF and we consider a wave function obtained from a restricted active space (RAS) calculation in this basis that would correspond to a CAS(8,26) calculation, using all the 26 orbitals originating from the $n=2$ and $n=3$ shells of the C atoms, but where only single (S), double (D), triple (T) and quadruple (Q) excitations are allowed in the active space. This wave function, that we denote by RAS-SDTQ(8,26), contains 110481 CSFs made of 411225 determinants.

The standard {\it ab initio} wave function is then multiplied by a Jastrow factor, imposing the electron-electron cusp conditions, but with essentially all other free parameters chosen to be $0$ to form our starting trial wave function. QMC calculations are performed with the program CHAMP~\cite{Cha-PROG-XX}, using this time the true Slater basis set rather than its Gaussian expansion. In comparison to GAMESS, additional symmetries outside the point group $D_{4h}$ are detected numerically which allows one to reduce the numbers of CSFs to 5, 50 and 165 for the CAS(8,5), CAS(8,7) and CAS(8,8) wave functions, respectively. For the large RAS-SDTQ(8,26) wave function, only a fraction of all the CSFs are retained in QMC by applying a variable cutoff on the CSF coefficients and an extrapolation procedure is used to estimate the QMC result if all the CSFs had been included (see Sec.~\ref{sec:systematic_improvement}).
For the orbital optimization, only the single excitations between orbitals of the same irreducible representation of $D_{\infty h}$ are generated. We, however, impose no restriction inside each of the two-dimensional irreducible representations $\pi_u$ and $\pi_g$. Although one can in principle identify the $\pi_x$ and $\pi_y$ components and forbid excitations between these two components to further reduce the number of free parameters, these redundancies appear to cause no problem in practice during the optimization. Also, we impose the electron-nucleus cusp condition on each orbital.
 The parameters of the trial wave function are optimized by the previously-described energy minimization procedures in VMC, using a very efficient accelerated Metropolis algorithm~\cite{Umr-PRL-93,Umr-INC-99}, allowing us to simultaneously make large Monte Carlo moves in configuration space and have a high acceptance probability. Once a trial wave function has been optimized, we perform a DMC calculation within the fixed-node and the short-time approximations (see, e.g., Refs.~\onlinecite{And-JCP-75,And-JCP-76,ReyCepAldLes-JCP-82,MosSchLeeKal-JCP-82}). We use an imaginary time step of $\tau=0.01$ H$^{-1}$ in an efficient DMC algorithm featuring very small time-step errors~\cite{UmrNigRun-JCP-93}, so that the accuracy is essentially limited by the quality of the nodal hypersurface of the trial wave function.

\section{Results and discussion}
\label{sec:results}

\subsection{Optimization of the Jastrow factor}

We first study the convergence behavior of the energy
minimization methods for the separate optimization of the Jastrow, CSF
and orbital parameters. To facilitate comparisons, we apply the VMC
optimization procedures with a common fixed statistical error of the
energy at each step, namely $0.5$~mHa. This is not the usual
way in which we routinely perform optimizations which is described
later in Sec.~\ref{sec:optall}.

\begin{figure}
\includegraphics[scale=0.35,angle=-90]{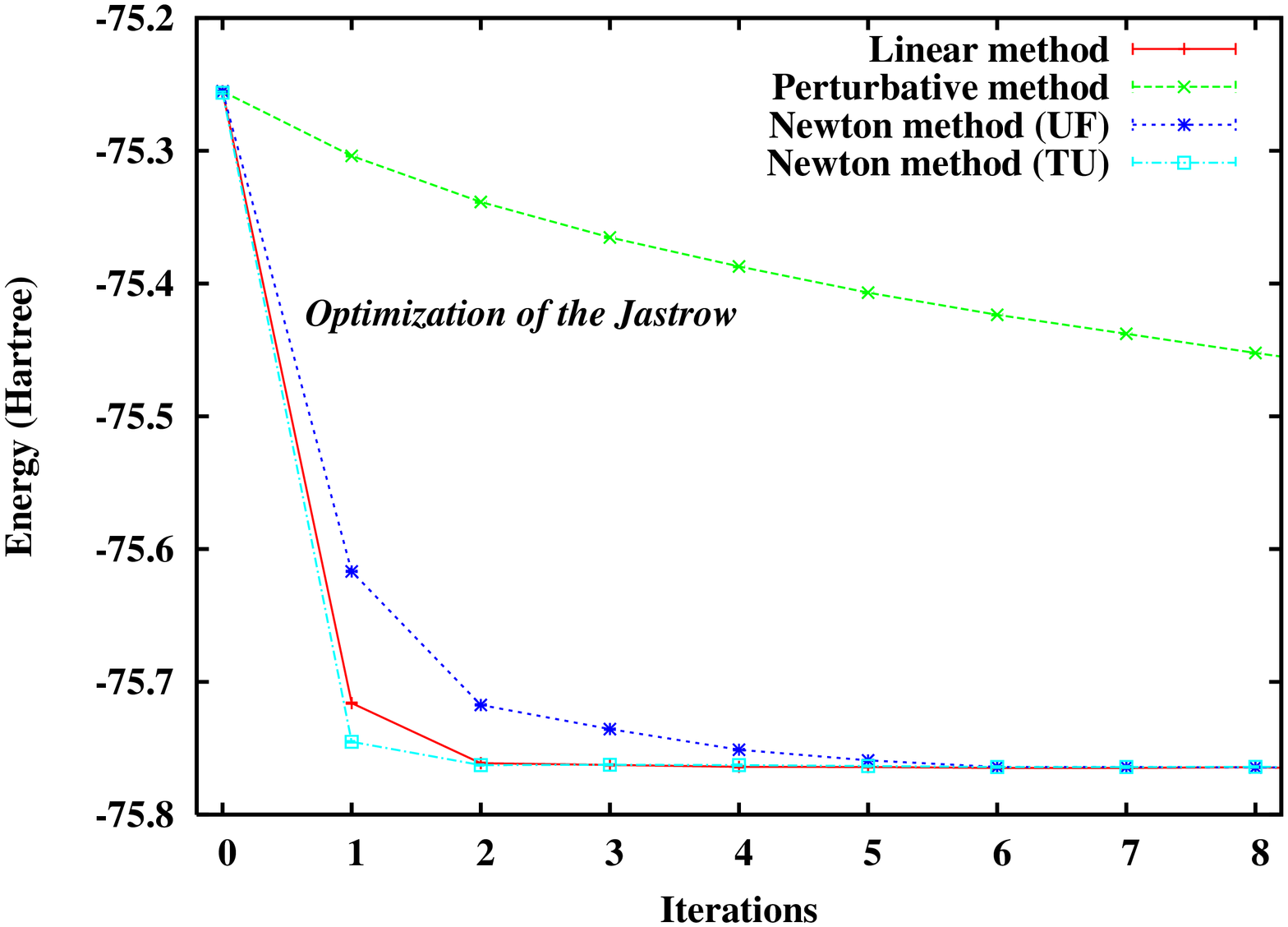}
\begin{picture}(0,0)(44,-160)
\includegraphics[scale=0.23,angle=-90]{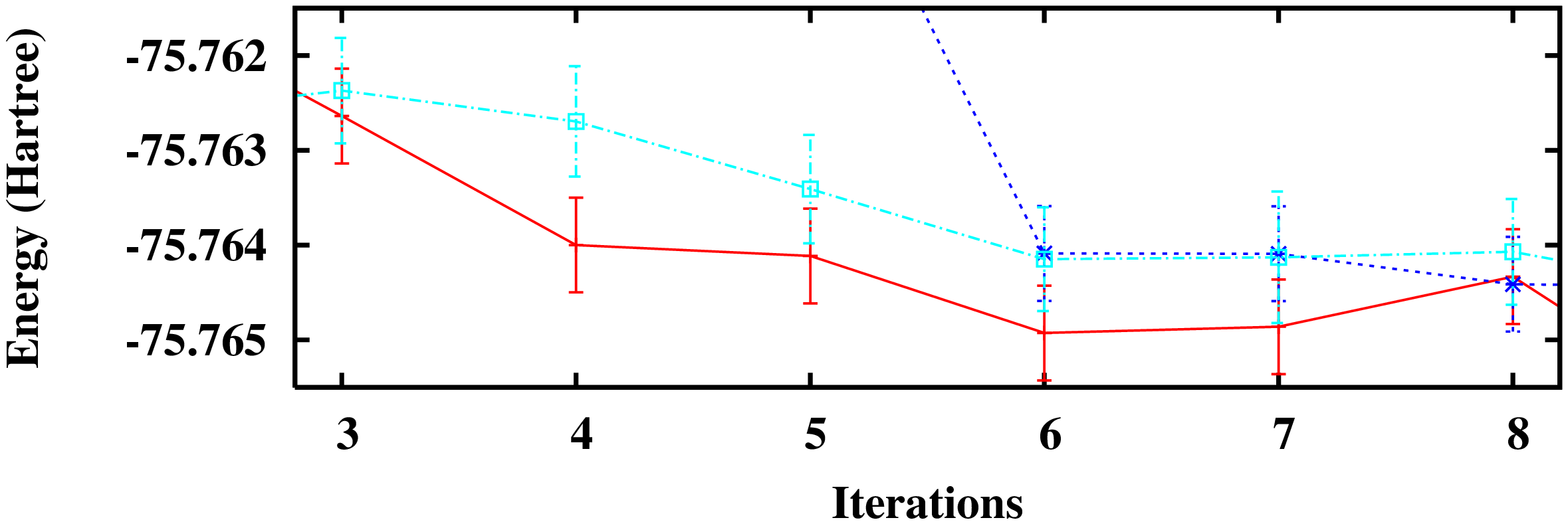}
\end{picture}
\caption{Convergence of the VMC total energy $E_\VMC$ of the all-electron C$_2$ molecule
during the optimization of the 24 Jastrow parameters in a wave function composed of the
RHF Slater determinant multiplied by a Jastrow factor. The linear, perturbative and
Newton energy minimization methods are compared. For the Newton method, the results
obtained with the UF Hessian of Eq.~(\ref{hijUF}) and the TU Hessian of Eq.~(\ref{hijTU})
are shown. The statistical error on the energy at each iteration is $0.5$~mHa.
The insert is an enlargement of the last 6 iterations.
}
\label{fig:c2_ktt_hfj_vmc_emin_j_opt_conv_stab}
\end{figure}

Figure~\ref{fig:c2_ktt_hfj_vmc_emin_j_opt_conv_stab} shows the
convergence of the total VMC energy during the optimization of the 24
Jastrow parameters in a wave function composed of the RHF Slater
determinant multiplied by a Jastrow factor. The linear, perturbative
and Newton methods are compared. For the Newton method, we present the
results obtained with the UF Hessian of Eq.~(\ref{hijUF}), already
used in Ref.~\onlinecite{UmrFil-PRL-05}, and with the TU Hessian of
Eq.~(\ref{hijTU}). To compare the fluctuations of these two Hessians,
we have computed the quantity $\eta = 1/N^\opt (N^\opt +1) \sum_{i=1}^{N^\opt} \sum_{j=i}^{N^\opt} (\sigma(h_{ij}))^2$
where $(\sigma(h_{ij}))^2$ is the variance of the element $h_{ij}$ of the Hessian averaged over 100 Monte Carlo configurations.
For the initial iteration of the optimization, far from the energy minimum, the UF Hessian fluctuates more than the TU Hessian
by a factor $\eta^{\text{UF}}/\eta^{\text{TU}}=3.6$.
For comparison, the LZR Hessian of Eq.~(\ref{hijLZR}) fluctuates more than the
TU Hessian by a factor $\eta^{\text{LZR}}/\eta^{\text{TU}}=150$,
more than two orders of magnitude larger even for this modest system.
Near the energy minimum, the factors are $\eta^{\text{UF}}/\eta^{\text{TU}}=3.3$
and $\eta^{\text{LZR}}/\eta^{\text{TU}}=600$.
These factors tend to increase with the system size.
The Newton method with the UF Hessian converges
reasonably fast in about 6 iterations, which is a little
faster than the convergence shown in Figs. 1,2 and 4 of
Ref.~\onlinecite{UmrFil-PRL-05} due to the previously-described correlated sampling
adjustment of the stabilizing constant $a_\diag$ in the course of the optimization
and despite the fact that we are performing an all-electron rather than a pseudopotential
calculation here\cite{UmrFil-PRL-05-note}.
The Newton method with the TU Hessian displays an even
faster convergence, the energy being essentially converged within the
statistical error at iteration 3 or 4. The linear method has a similar
convergence rate to the Newton method with the TU Hessian.
The Newton method with the TU Hessian and the linear method are both
stable even without stabilization if sufficiently large Monte Carlo samples are used.
When stabilization is employed, the stabilization constant $a_\diag$ remains
small during the optimization, in this example from $10^{-3}$ for the
initial iteration to $10^{-7}$ for the last iterations which is 2 or 3
orders of magnitude smaller than the values of $a_\diag$ in
the Newton method with the UF Hessian. The perturbative method, in
contrast, converges very slowly. In fact, it turns out that the energy
denominators for the Jastrow parameters, $\Delta {\cal E}_{\alpha_i}$,
calculated according to Eq.~(\ref{DeltaEi}), are all of order unity
and $a_\diag$ needs to be increased to as much as $10^2$ to retain
stability. In this case, the perturbative method essentially reduces
to the inefficient SR optimization technique.

\subsection{Optimization of the CSF coefficients}

\begin{figure}
\includegraphics[scale=0.35,angle=-90]{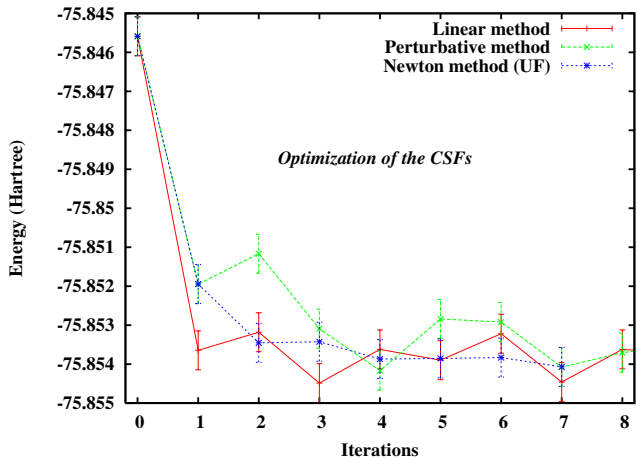}
\caption{Convergence of the VMC total energy $E_\VMC$ of the all-electron C$_2$ molecule during the optimization of the 49 CSF parameters in a wave function composed of a CAS(8,7) part multiplied by a previously-optimized Jastrow factor. The linear, perturbative and Newton [with the UF Hessian of Eq.~(\ref{hijUF})] energy minimization methods are compared. The statistical error on the energy at each iteration is $0.5$~mHa.
}
\label{fig:c2_ktt_cas87j_vmc_emin_csf_opt_conv_stab}
\end{figure}

Figure~\ref{fig:c2_ktt_cas87j_vmc_emin_csf_opt_conv_stab} shows the
convergence of the total VMC energy during the optimization of the 49
CSF parameters in a wave function composed of a CAS(8,7) determinantal
part multiplied by a previously-optimized Jastrow factor, using the
linear, perturbative and Newton [with the UF Hessian of
Eq.~(\ref{hijUF})] methods. The linear method converges in $1$
iteration, as it must, and does not require any stabilization.
When stabilization is used, $a_\diag$
remains as low as $10^{-6}$ to $10^{-8}$ during the whole optimization.
The Newton and perturbative methods
converge in $2$ or $3$ iterations, and are not as intrinsically stable, $a_\diag$ being a few
orders of magnitude larger for the Newton method
and several orders of magnitude larger for the perturbative method.
The energy denominators for
the CSF parameters in the perturbative method, $\Delta {\cal
  E}_{c_I}$, calculated according to Eq.~(\ref{DeltaEi}),
span only one order of magnitude.

\subsection{Optimization of the orbitals}

\begin{figure}
\includegraphics[scale=0.35,angle=-90]{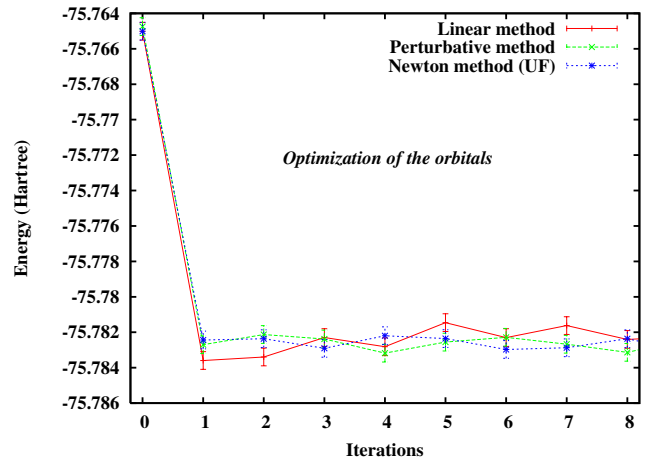}
\caption{Convergence of the VMC total energy $E_\VMC$ of the all-electron C$_2$ molecule during the optimization of the 44 orbital parameters in a wave function composed of a single Slater determinant multiplied by a previously-optimized Jastrow factor. The linear, perturbative and Newton [with the UF Hessian of Eq.~(\ref{hijUF})] energy minimization methods are compared. The statistical error on the energy at each iteration is $0.5$~mHa.
}
\label{fig:c2_ktt_hfj_vmc_emin_o_opt_conv_stab}
\end{figure}

Figure~\ref{fig:c2_ktt_hfj_vmc_emin_o_opt_conv_stab} shows the
convergence of the total VMC energy during the optimization of all the
44 orbital parameters in a wave function composed of a single Slater
determinant multiplied by a previously-optimized Jastrow factor, using
the linear, perturbative and Newton [with the UF Hessian of
Eq.~(\ref{hijUF})] methods. The three methods display very similar
convergence rates, the energy being converged within the statistical
error in $1$ iteration using any of the three methods.
In this example, the linear and perturbative methods converged even without
stabilization whereas the Newton method required stabilization.
The energy denominators for the orbital parameters in the perturbative method,
$\Delta {\cal E}_{kl}$, calculated according to Eq.~(\ref{DeltaEi}),
typically span two orders of magnitude from $1$ to $100$.

\begin{figure}
\includegraphics[scale=0.35,angle=-90]{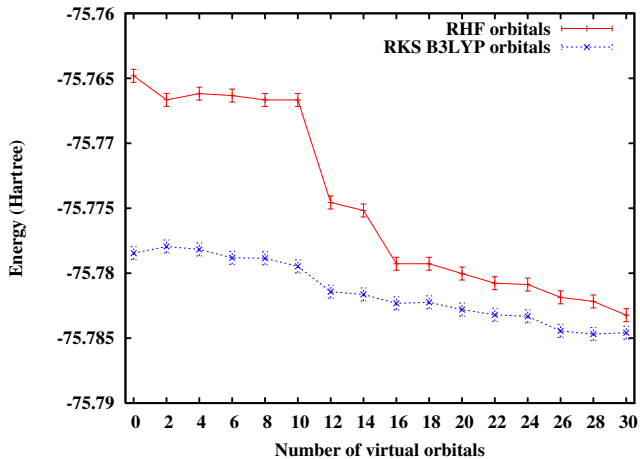}
\caption{Total VMC energy $E_\VMC$ of the all-electron C$_2$ molecule with respect to the number of virtual orbitals included in the optimization of the orbital parameters in a wave function composed of a single Slater determinant multiplied by a previously-optimized Jastrow factor, using RHF and RKS B3LYP starting orbitals.
The orbitals are ordered according to their energies.
The statistical error on the energy is $0.5$~mHa.
}
\label{fig:c2_ktt_hfj_vmc_emin_o_opt3_orb}
\end{figure}

\begin{figure*}
\includegraphics[scale=0.35,angle=-90]{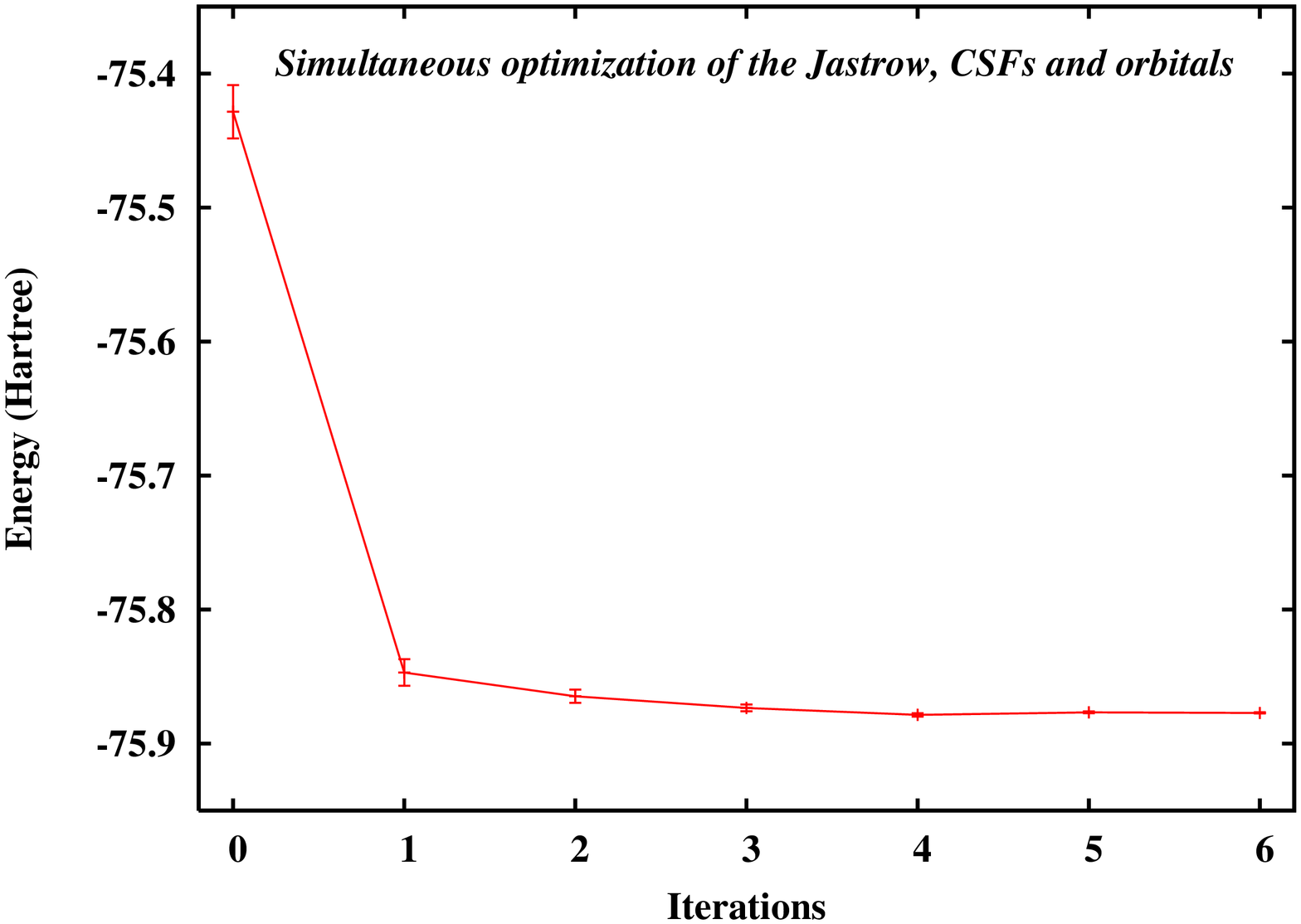}
\includegraphics[scale=0.35,angle=-90]{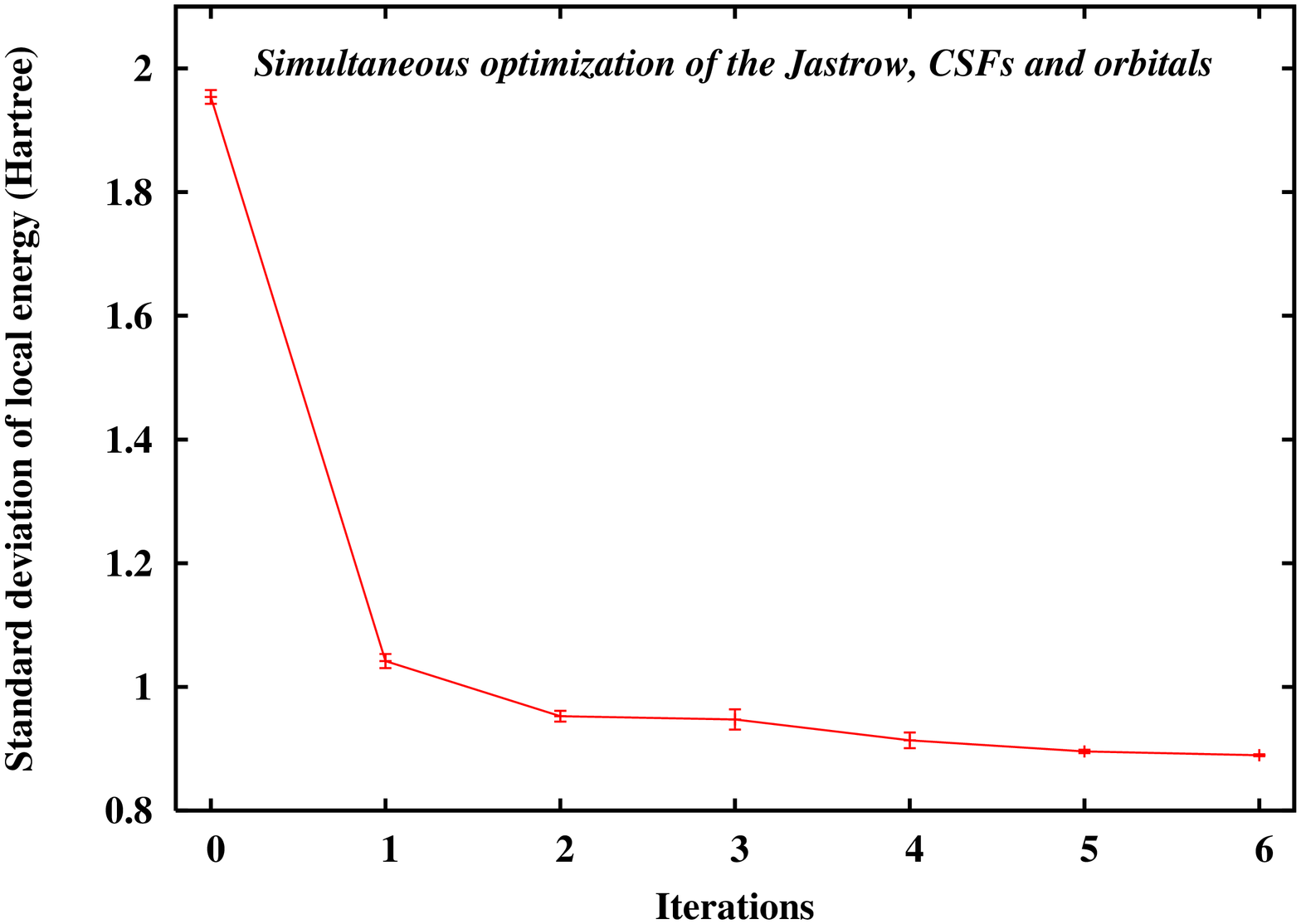}
\begin{picture}(0,0)(183,-175)
\includegraphics[scale=0.24,angle=-90]{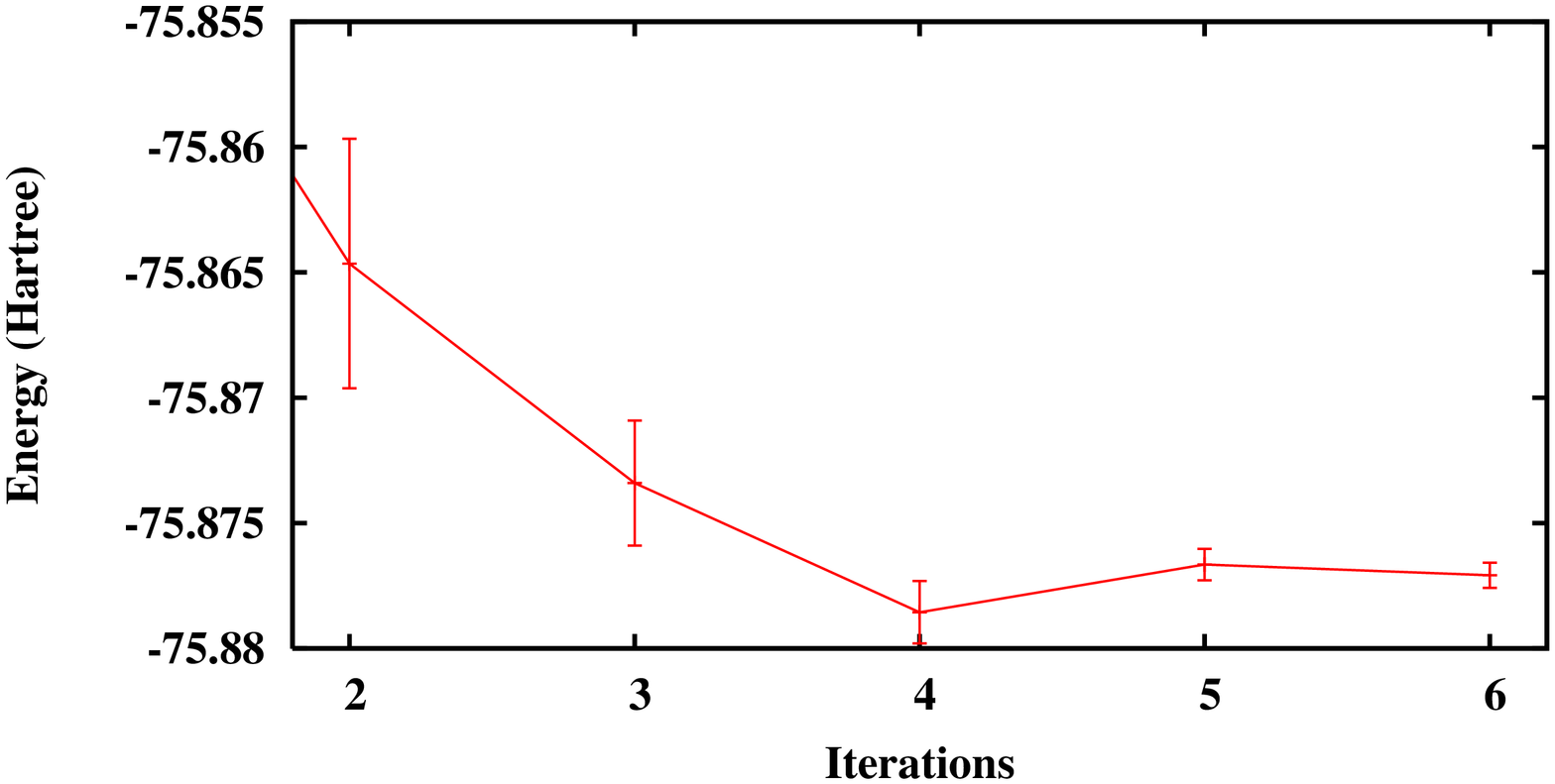}
\end{picture}
\begin{picture}(0,0)(-72,-175)
\includegraphics[scale=0.24,angle=-90]{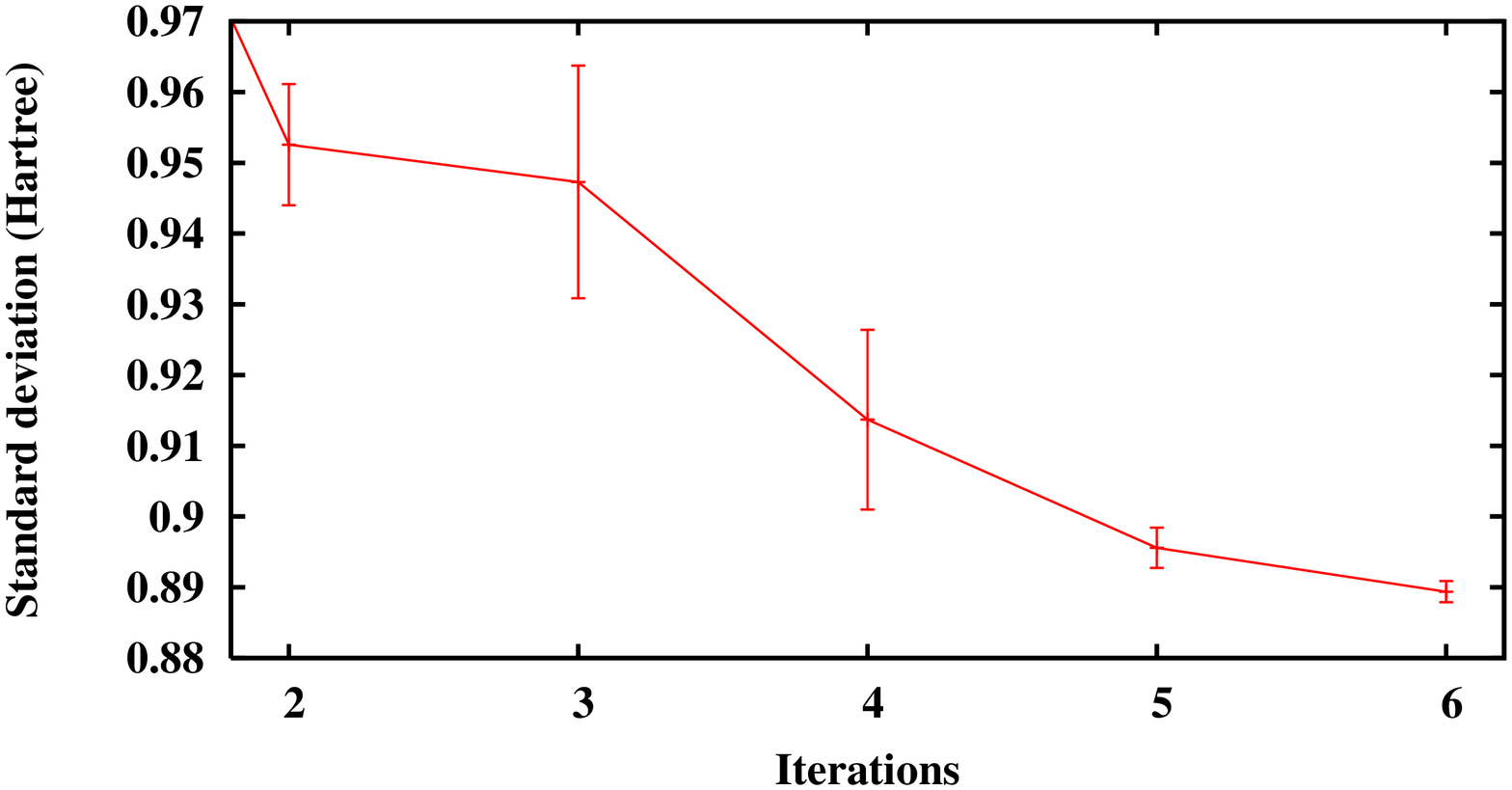}
\end{picture}
\caption{Convergence of the VMC total energy $E_\VMC$ (left plot) and of the VMC standard deviation of the local energy $\sigma_\VMC$ (right plot) of the all-electron C$_2$ molecule during the \textit{simultaneous} optimization of the 24 Jastrow parameters, 49 CSF parameters and 64 orbital parameters in a wave function composed of the CAS(8,7) determinantal part multiplied by a Jastrow factor, using the linear energy minimization method. The statistical error on the energy is initially of $0.02$~Ha and is decreased by a factor 2 at each iteration until $0.5$~mHa. The insets are enlargements of the last 5 iterations.
}
\label{fig:c2_ktt_cas87j_vmc_emin_csfoj_opt3_lin_new2}
\end{figure*}

In the previous orbital optimization, we have considered a full
optimization of all the orbital parameters, i.e. all the allowed
excitations from the 6 closed occupied orbitals to the 30 virtual
orbitals were included in the calculation. One may also consider a
partial orbital optimization by restricting the excitations to the
lowest several virtual orbitals, as also proposed within the EFP or
perturbative EFP approaches~\cite{Fil-PRIV-XX}. This allows one to
reduce the computational effort and also to decrease the statistical
noise in the calculation since it is the excitations to the
highest-lying virtual orbitals that modify the most the nodal
structure of the wave function, leading to large fluctuations of the
ratio $\Psi_i(\b{R})/\Psi_0(\b{R})$.
Fig.~\ref{fig:c2_ktt_hfj_vmc_emin_o_opt3_orb} shows the total VMC
energy with respect to the number of virtual orbitals included in the
optimization for a wave function composed of a single Slater
determinant multiplied by a previously-optimized Jastrow factor. Two
sets of starting orbitals are compared: orbitals obtained from a RHF
calculation and orbitals obtained from a restricted Kohn-Sham (RKS)
calculation with the hybrid exchange-correlation functional
B3LYP~\cite{Bec-JCP-93,SteDevChaFri-JPC-94}, using the ordering given
by the orbital energies. In both cases, as expected, the energy
decreases monotonically within the statistical error as the number of virtual
orbitals included in the optimization increases.
However the slope of the energy does not change monotonically and
it is necessary to include almost all the orbitals to get close to the optimal energy.
From Fig.~\ref{fig:c2_ktt_hfj_vmc_emin_o_opt3_orb} we see that for the C$_2$ molecule
the B3LYP orbitals provide a better starting point than the RHF orbitals.
In our experience, this is often but not always the case.
It is possible that the selection of the virtual orbitals adopted here,
based on the orbital energy ordering, may not be the best choice and other
selections based on symmetry or chemical intuition could lead to a
more rapid convergence.

Note that Fig.~\ref{fig:c2_ktt_hfj_vmc_emin_o_opt3_orb} was obtained
by just optimizing the orbital parameters for a fixed, previously-optimized Jastrow factor.
If instead the Jastrow and orbital parameters are optimized simultaneously a significantly lower energy
is obtained, e.g. including all 30 virtual orbitals gives an energy of -75.8069(5)~Ha (see
Table~\ref{tab:energies}) as opposed to -75.7845(5)~Ha in Fig.~\ref{fig:c2_ktt_hfj_vmc_emin_o_opt3_orb}.

To summarize, the Newton and the linear methods converge very rapidly when
optimizing any kind of parameter,
though the linear method is more stable for the optimization of the
determinantal part of the wave function. The perturbative method is a
good, less expensive alternative for the optimization of the orbital
parameters and, to a lesser extent, for the optimization of the CSF
parameters, but is very slowly convergent for the Jastrow parameters.

It is clear from Eq.~(\ref{Deltaalpha1}) that the perturbative method
can be viewed as a Newton method with an approximate Hessian.
The poor behavior of the perturbative method for the Jastrow parameters
means that this Hessian is a bad approximation to the exact Hessian, whose eigenvalues
span more than 10 orders of magnitude for these parameters.
In fact, any method based on an approximate Hessian that is not able to
reproduce all these orders of magnitude, such as the steepest-descent method,
 is bound to converge very slowly.
On the other hand, the eigenvalues of the Hessian for the CSF and orbital parameters
span only a couple of orders of magnitude and the approximate Hessian of the perturbative method is
sufficient to allow rapid convergence.

\subsection{Optimization of all the parameters: simultaneous or alternated optimization?}
\label{sec:optall}

After having studied the behavior of the energy minimization methods
for the optimization of each kind of parameter, we now move on to the
more practical problem of how to optimize all the parameters.

The most obvious possibility is to optimize \textit{simultaneously}
the Jastrow, CSF and orbital parameters using the linear method, the
method having the best overall efficiency for all these parameters. In
practice, we proceed as follows. We start an optimization run with a
short Monte Carlo simulation with a large statistical error (e.g.,
$0.02$~Ha for the C$_2$ molecule), and we decrease progressively the
statistical error at each iteration until the energy is converged to
$10^{-3}$~Ha for three consecutive iterations.  We choose the optimal
parameters to be those from the iteration with the smallest value of
$E_\VMC$ plus three times the statistical error of $E_\VMC$, which is
often but not always the last iteration.
A typical example of the convergence of the total VMC
energy and of the standard deviation $\sigma_\VMC$ is shown in
Fig.~\ref{fig:c2_ktt_cas87j_vmc_emin_csfoj_opt3_lin_new2} for the
simultaneous optimization of the Jastrow, CSF and orbital parameters
in a wave function composed of a CAS(8,7) determinantal part
multiplied by a Jastrow factor. In this case, the energy converges in
4 or 5 iterations. The standard deviation typically converges a little
slower than the energy since we are optimizing just the energy here.
A faster convergence, to a somewhat smaller value of the standard deviation,
can be achieved by optimizing a linear combination of the energy and variance as in
Ref.~\cite{UmrFil-PRL-05}.

\begin{figure*}
\includegraphics[scale=0.35,angle=-90]{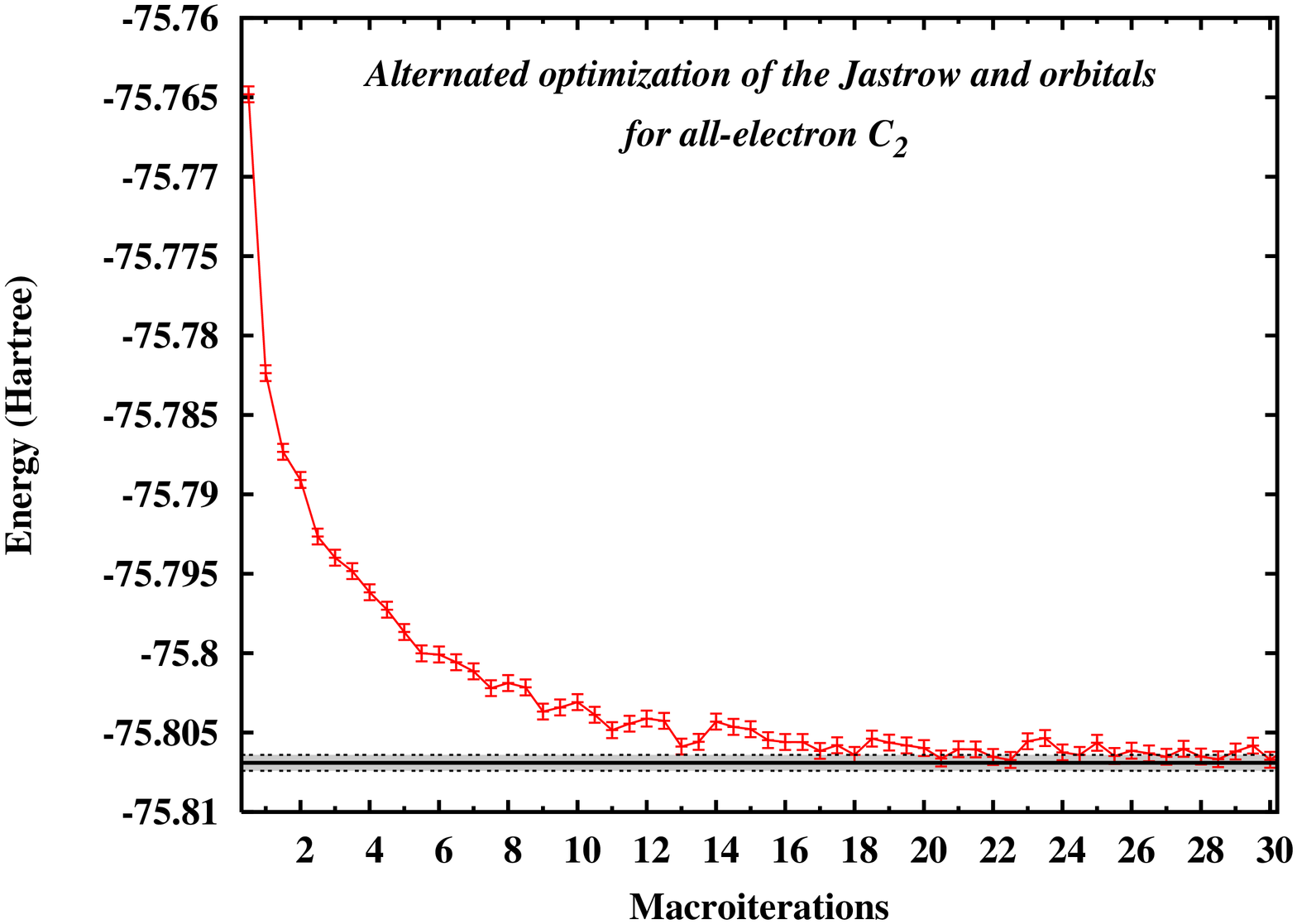}
\includegraphics[scale=0.35,angle=-90]{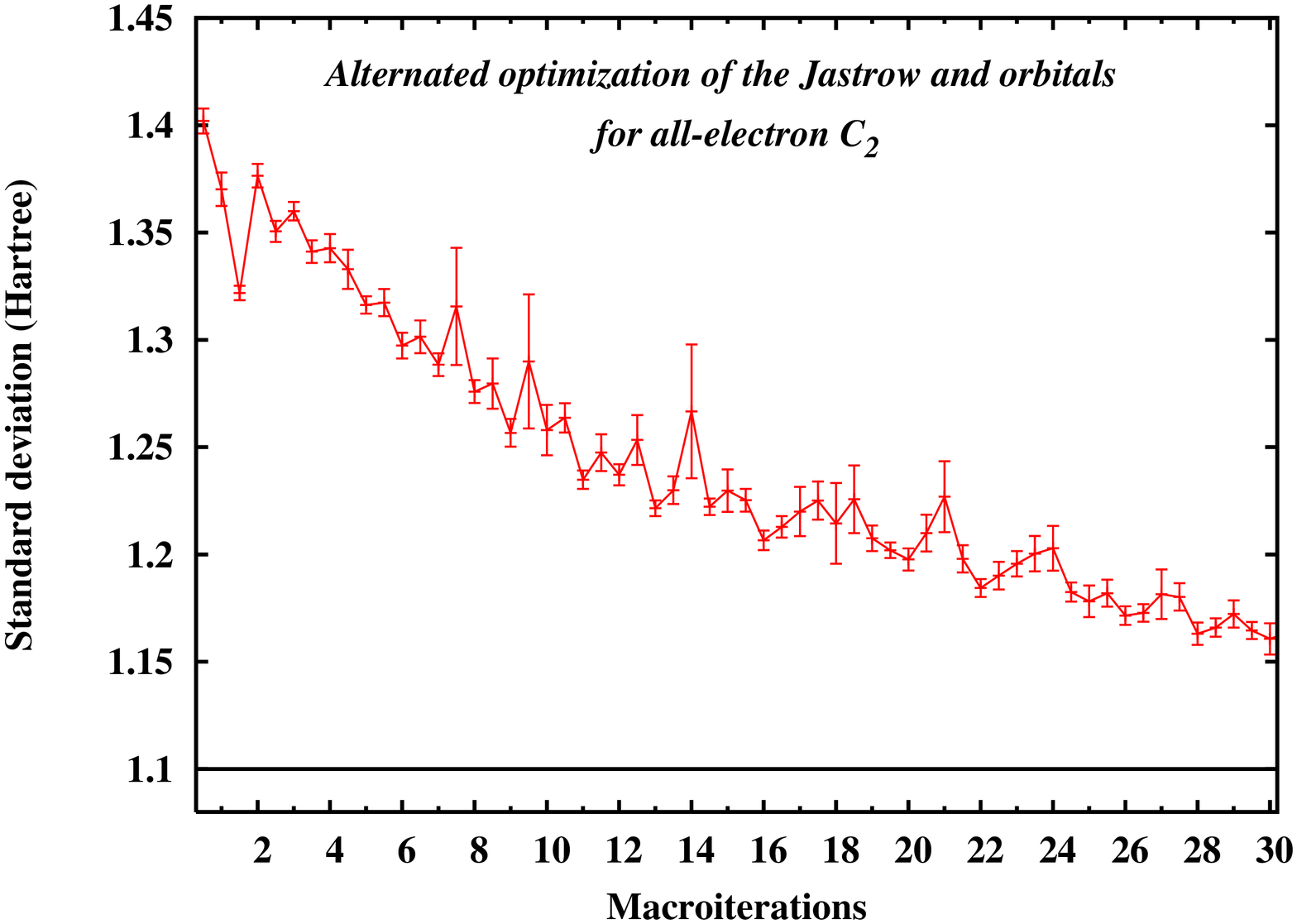}
\caption{Demonstration of the slow convergence of the VMC total energy $E_\VMC$ (left plot) and of
the VMC standard deviation of the local energy $\sigma_\VMC$ (right plot)
of the all-electron C$_2$ molecule during the \textit{alternated}
optimization of the 24 Jastrow parameters and 44 orbital parameters in a
wave function composed of a single Slater determinant multiplied by a Jastrow factor.
The half-integer macroiteration numbers correspond to the optimization of
the Jastrow factor and the integer macroiteration numbers correspond to
the optimization of the orbitals. The statistical error on the energy is always $0.5$~mHa.
The simultaneous optimization of the Jastrow and orbital parameters gives
an energy of $-75.8069(5)$~Ha and a standard deviation of $1.1$~Ha, indicated on the plots by horizontal lines.
}
\label{fig:c2_ktt_hfj_vmc_emin_oj_lin_alter}
\end{figure*}

\begin{figure}
\includegraphics[scale=0.35,angle=-90]{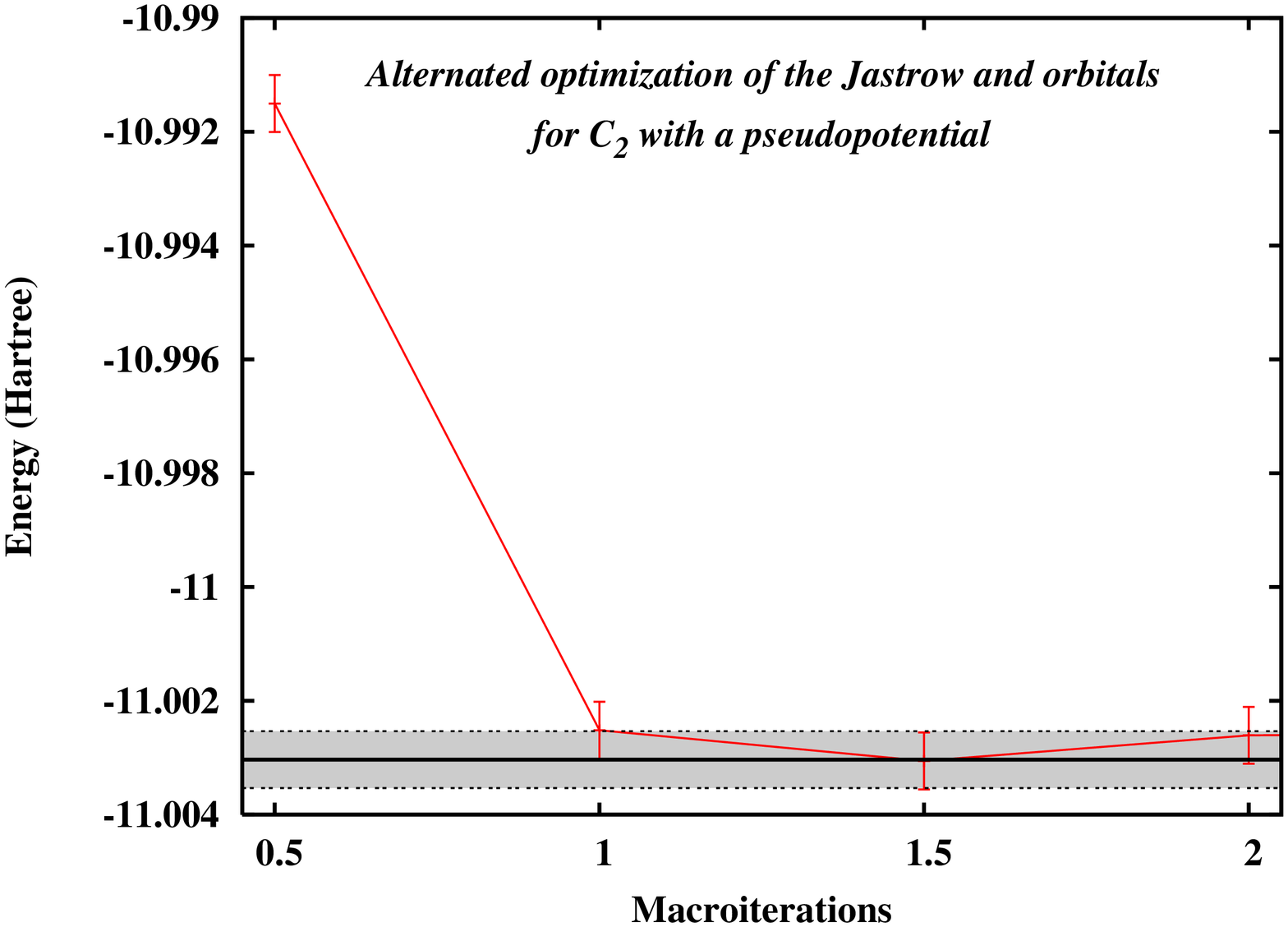}
\caption{Convergence of the VMC total energy $E_\VMC$ of the C$_2$ molecule with a pseudopotential removing the core electrons during the \textit{alternated} optimization of the 24 Jastrow parameters and 42 orbital parameters in a wave function composed of a single Slater determinant multiplied by a Jastrow factor. The half-integer macroiteration numbers correspond to the optimization of the Jastrow factor and the integer macroiteration numbers correspond to the optimization of the orbitals. The statistical error on the energy is always $0.5$~mHa. The \textit{simultaneous} optimization of the Jastrow and orbital parameters gives an energy of $-11.0030(5)$~Ha, indicated on the plot by horizontal lines.
}
\label{fig:c2_ps_hfj_vmc_emin_oj_lin_alter}
\end{figure}

Another possibility is to \textit{alternate} between the optimization
of the different kinds of parameters until global convergence. This
has the advantage of allowing one to use different optimization
methods for the various parameters, e.g. optimization of the Jastrow
factor and the CSF coefficients with the Newton or linear method and
optimization of the orbitals with the less expensive but still very
efficient perturbative method.
Fig.~\ref{fig:c2_ktt_hfj_vmc_emin_oj_lin_alter} shows the convergence
of the total VMC energy and of the standard deviation during the
alternated optimization of the Jastrow parameters and of the orbital
parameters in a wave function composed of a single Slater determinant
multiplied by a Jastrow factor for the all-electron C$_2$ molecule.
The convergence of the energy is surprisingly very slow, the
convergence of the standard deviation is even worse. This is an indication
of the presence of a strong coupling between some Jastrow and orbital parameters.
This situation is in sharp contrast with the case where a pseudopotential is used
to remove the core electrons.
Fig.~\ref{fig:c2_ps_hfj_vmc_emin_oj_lin_alter} shows the convergence
of the total VMC energy during the alternated optimization of the
Jastrow parameters and of the orbital parameters in a wave function
composed of a single Slater determinant multiplied by a Jastrow factor
for the C$_2$ molecule with a Hartree-Fock
pseudopotential~\cite{Shi-UNP-XX} and an adequate Gaussian
one-electron basis set. The convergence is very fast, the energy being
essentially converged within the statistical error in $1$~macroiteration.
This favorable behavior has already been observed in
other systems with pseudopotentials~\cite{Fil-PRIV-XX}, but we have
also found pseudopotential systems for which the convergence is not as fast.

For the all-electron case, it thus seems that simultaneous optimization of the parameters is much preferable.
The coupling  between the different parameters seems to be too strong to allow
an efficient alternated optimization.
For large systems most of the wave function parameters are orbital and CSF parameters
for which the perturbative method works well.
It seems then promising to simultaneously optimize all the parameters
with the Newton or the linear methods, using for the part of the Hessian or the Hamiltonian
matrices involving the CSF and orbital coefficients rough approximations
inspired by the perturbative method~\cite{FilTouUmr-JJJ-XX}.

\subsection{Systematic improvement by wave function optimization}
\label{sec:systematic_improvement}

\begin{figure}
\vspace{0.5cm}
\includegraphics[scale=0.35,angle=-90]{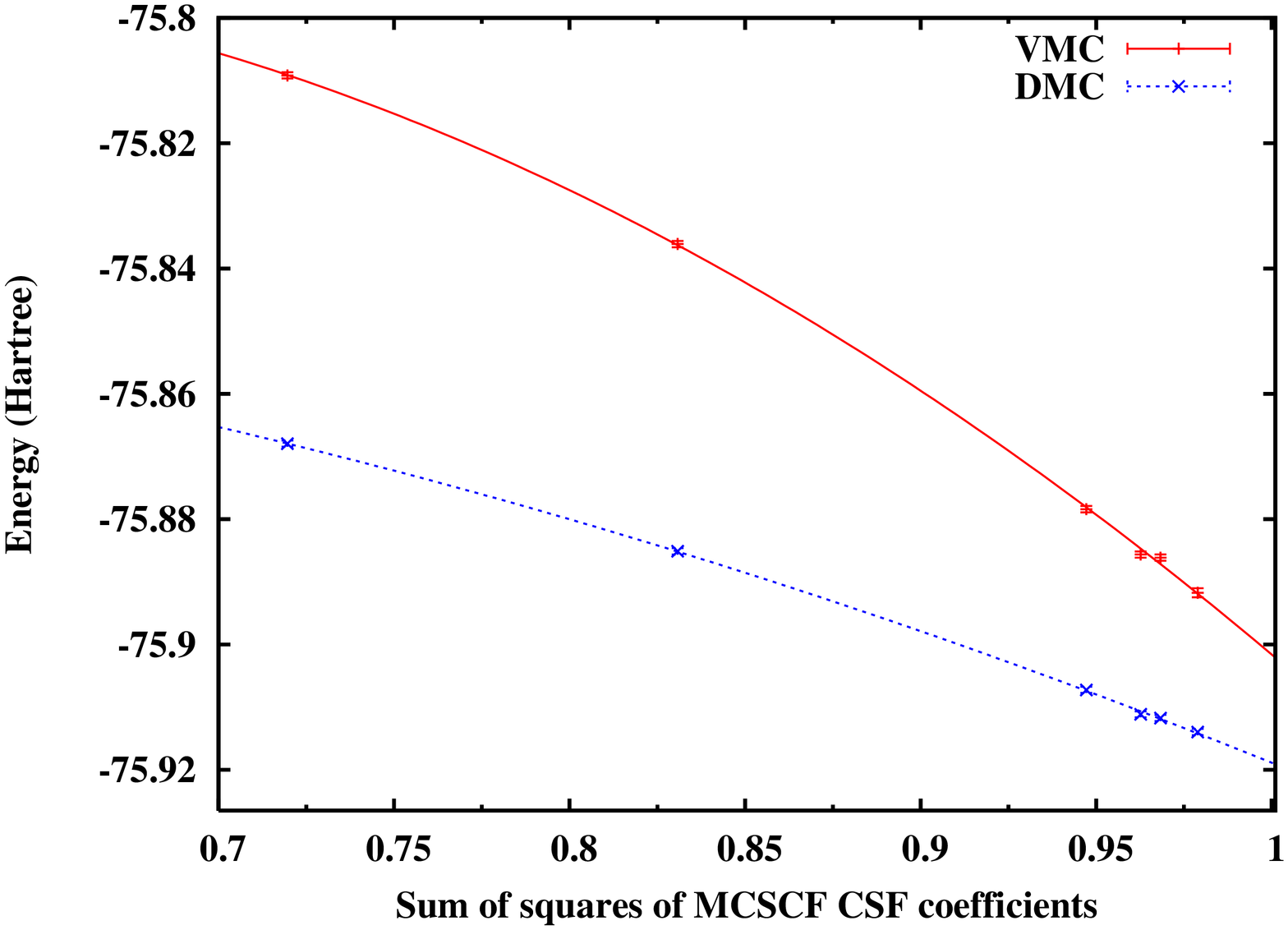}
\caption{VMC and DMC energies obtained with the truncated, fully reoptimized Jastrow-Slater RAS-SDTQ(8,26) wave functions with respect to the sum of the squares of the MCSCF CSF coefficients retained, $\sum_{i=1}^{N_\CSF} (c_i^{\text{MCSCF}})^2$. The latter quantity is equal to $1$ in the limit where all the CSFs of the RAS-SDTQ(8,26) calculation are kept in the wave function which is extrapolated by quadratic fits.
}
\label{fig:c2_6s4p1d_csfj_cusp_emin_csfoj}
\vspace{0.8cm}
\end{figure}

\begin{table*}[t]
\begin{tabular}[b]{l |l |c c c }

\hline\hline
\multicolumn{1}{c|}{Wave function form} & \multicolumn{1}{c|}{Parameters optimized in VMC} & $E_{\VMC}$   & $ E_{\DMC}$  & \multicolumn{1}{c}{$\sigma_{\VMC}$} \\
\hline

Jastrow $\times$ determinant & Jastrow (24)            & -75.7648(5)  & -75.8570(5)  & 1.4\\
                             & Jastrow (24) + orbitals (44) & -75.8069(5)  & -75.8682(5)  & 1.1 \\
\hline
Jastrow $\times$ CAS(8,5) & Jastrow (24)                   & -75.8045(3)  & -75.8750(5)  & 1.3 \\
                          & Jastrow (24) + CSFs (6)            & -75.8094(5)  & -75.8807(5)  & 1.3 \\
                          & Jastrow (24) + CSFs (6)+ orbitals (52) & -75.8374(5)  & -75.8882(5)  & 1.0 \\
\hline
Jastrow $\times$ CAS(8,7) & Jastrow (24)                   & -75.8469(5)  & -75.8973(5)  & 1.2 \\
                          & Jastrow (24) + CSFs (49)            & -75.8546(5)  & -75.9032(5)  & 1.2 \\
                          & Jastrow (24) + CSFs (49) + orbitals (64) & -75.8769(5)  & -75.9092(5)  & 0.9 \\

\hline
Jastrow $\times$ CAS(8,8) & Jastrow (24)                   & -75.8462(5)     &  -75.8999(6)   & 1.1 \\
                          & Jastrow (24) + CSFs (164)            & -75.8562(5)     &  -75.9050(6)    &  1.1 \\
                          & Jastrow (24) + CSFs (164) + orbitals (70) & -75.8801(6)     &  -75.9099(5)   & 0.9 \\
\hline
Jastrow $\times$ RAS-SDTQ(8,26) & Jastrow + CSFs + orbitals (extrapolation) & -75.9016(5) & -75.9191(5) & -- \\
\hline
\multicolumn{2}{l|}{Exact} &  &   -75.9265(8) & \\
\hline\hline

\end{tabular}
\caption{Total VMC and DMC energies, $E_\VMC$ and $E_\DMC$, and VMC standard deviation of
the local energy $\sigma_\VMC$ of the C$_2$ molecule for different trial wave functions
and different levels of optimization. The kind and number of optimized parameters are indicated.
When not optimized in VMC, the CSF and orbital coefficients have been fixed at their RHF values for the single-determinant case and
 at their CAS MCSCF values for the multiconfiguration cases.
For the large Jastrow $\times$ RAS-SDTQ(8,26) wave function, the VMC and DMC values are obtained by an extrapolation procedure (see Sec.~\ref{sec:systematic_improvement} and Fig.~\ref{fig:c2_6s4p1d_csfj_cusp_emin_csfoj}).
For the energies, the numbers in parentheses are estimates of the statistical error on the last digit. All units are Hartree.
}
\label{tab:energies}
\end{table*}

Table~\ref{tab:energies} reports the total VMC and DMC energies, $E_\VMC$ and
$E_\DMC$, and the VMC standard deviation of the local
energy $\sigma_\VMC = \sqrt{\langle E_\L(\b{R})^2\rangle - \langle
 E_\L(\b{R})\rangle^2}$ for the different trial wave
functions considered. For the single-determinant, CAS(8,5), CAS(8,7) and CAS(8,8) wave functions,
we present the results for three levels of optimization. At the first
level, only the Jastrow factor is optimized. At the second level, the
Jastrow factor and the CSF coefficients are optimized together. At the
third level, the Jastrow factor, the CSF coefficients and the orbitals
are all optimized together. Going from one level to the next one
improves the accuracy of the wave function but also increases the computational
cost of the optimization.
We note that it is important to reoptimize the determinantal (CSF and orbital) parameters,
along with the Jastrow parameters, rather than keeping them fixed at the values obtained
from the MCSCF wave functions.
For each wave function, the effect of
reoptimizing the determinantal part is to lower the VMC energy by
about $0.03$ to $0.04$~Ha, and the standard deviation of the energy by about $0.2$ to
$0.3$~Ha. More remarkably, even though the optimization is performed at the
VMC level, the DMC energy also goes down by about $0.01$~Ha,
implying that the nodal hypersurface of the trial wave function also
improves. In addition, one observes a systematic improvement of the
VMC and DMC energies when the size of the CAS increases, provided that at
least the CSF coefficients are reoptimized with the Jastrow factor.

Including all the $110481$ CSFs of the RAS-SDTQ(8,26) wave function is too costly in
quantum Monte Carlo but one can use a series of truncated wave functions obtained by
retaining only small numbers of CSFs with coefficients larger in absolute value than
a variable cutoff, and then estimate the energy by extrapolation to the
limit that all the CSFs are kept. Fig.~\ref{fig:c2_6s4p1d_csfj_cusp_emin_csfoj} shows
the VMC and DMC energies obtained with these truncated, fully reoptimized multi-determinantal
wave functions with respect to the sum of the squares of the MCSCF CSF coefficients retained,
$\sum_{i=1}^{N_\CSF} (c_i^{\text{MCSCF}})^2$.
Since the RAS-SDTQ(8,26) wave function is normalized, the latter quantity is equal to $1$
in the limit where all the CSFs are kept in the wave function.
Experience shows that the energies are well extrapolated by quadratic fits. The extrapolated
DMC energy is $-75.9191(5)$ which amounts for $98.6\%$ of the correlation energy
(using the HF energy of -75.40620~Ha calculated in Ref.~\onlinecite{CadWah-ADNDT-74}).

On the other hand, to calculate accurate well depths (dissociation energy + zero-point energy) it is often sufficient to rely on some partial cancellation of error between the atom and the molecule by employing atomic and molecular wave functions that are consistent with each other. For example, using the DMC energy of the C$_2$ molecule given by the Jastrow-Slater full-valence CAS(8,8) wave function and the DMC energy of the C atom given by the consistent Jastrow-Slater full-valence CAS(4,4) wave function with the same one-electron basis leads to a well depth of 6.46(1) eV, in perfect agreement within the uncertainty with the exact, nonrelativistic well depth estimated at 6.44(2) eV~\cite{BytRue-JCP-05,BytRue-JCP-05-note}. In contrast, the well depth calculated from MCSCF with the molecular CAS(8,8) and atomic CAS(4,4) wave functions (without Jastrow factor) is 5.62 eV, in poor agreement with the exact value.

\section{Conclusions}
\label{sec:conclusion}

We have studied three wave function optimization methods based on
energy minimization in a VMC context: the Newton, linear and
perturbative methods. These general methods have been applied here to
the optimization of wave functions consisting of a multiconfiguration
expansion multiplied by a Jastrow factor for the all-electron C$_2$
molecule.
The Newton and linear methods are both very efficient for the optimization
of the Jastrow, CSF and orbital parameters, the linear method being generally more stable.
The less computationally expensive perturbative method is efficient only for the CSF and orbital parameters.
We have used the linear method to simultaneously optimize the Jastrow,
CSF and orbital parameters, a much more efficient procedure than alternating
between optimizing the different kinds of parameters.
The linear method is capable of yielding not only ground state energies
but excited state energies as well~\cite{NigMel-PRL-01}.

Although the optimization is performed at the VMC level, we have
observed for the C$_2$ molecule studied here, as well as for other systems
not discussed in the present paper, that as more
parameters are optimized the DMC energies decrease monotonically,
implying that the nodal hypersurface also improves monotonically.
In fact, a sequence of trial wave
functions consisting of multiconfiguration expansions of increasing sizes
multiplied by a Jastrow factor, with all the Jastrow, CSF and orbital
parameters optimized together allows one to systematically reduce the
fixed-node error of DMC calculations for the systems studied.

Future directions for this work include optimization of the exponents
of the one-electron basis functions (either Slater or Gaussian functions),
direct optimization of the DMC energy and optimization of the geometry.

\begin{acknowledgments}
  We thank Claudia Filippi, Peter Nightingale, Sandro Sorella, Richard Hennig, Roland Assaraf, Andreas Savin,
  Anthony Scemama, Wissam Al-Saidi and Paola Gori-Giorgi for
  stimulating discussions and useful comments on the manuscript,
  and Eric Shirley for having provided us with the code for
  generating Hartree-Fock pseudopotentials. This work was supported in
  part by the National Science Foundation (DMR-0205328, EAR-0530301), Sandia National Laboratory,
  a Marie Curie Outgoing International Fellowship (039750-QMC-DFT),
  and DOE-CMSN.
  The calculations were performed at the Cornell Nanoscale Facility and the Theory Center.
\end{acknowledgments}

\bibliographystyle{apsrev}
\bibliography{biblio}

\begin{thebibliography}{61}
\bibitem[{Umr()}]{UmrFil-PRL-05-note}
\bibinfo{note}{The convergence in 2 iterations, mentioned near the end of
  Ref.~\onlinecite{UmrFil-PRL-05} was obtained using the correlated sampling
  adjustment of $a_\diag$ and the TU Hessian.}
\bibitem[{Byt()}]{BytRue-JCP-05-note}
\bibinfo{note}{Note that this estimate of the exact well depth differs from the
  one used in Ref.~\onlinecite{UmrTouFilSorHeg-JJJ-XX} where we used instead
  the scalar-relativistic, valence-corrected estimate of
  Ref.~\onlinecite{BytRue-JCP-05} since calculations were performed with a
  relativistic pseudopotential.}
\end{thebibliography}

\end{document}